\documentstyle[12pt,axodraw]{article}
\newcommand{\JHEP}{J. High Energy Phys. }
\newcommand{\RMP}{Rev. Mod. Phys. }
\newcommand{\NP}{Nucl. Phys. }

\newcommand{\PRL}{Phys. Rev. Lett. }
\newcommand{\PL}{Phys. Lett. }

\newcommand{\gmunu}{g_{\mu\nu}}
\newcommand{\Aalf}{A_{\alpha}}\newcommand{\Abet}{A_{\beta}}
\newcommand{\Agam}{A_{\gamma}}
\newcommand{\Zalf}{Z_{\alpha}}\newcommand{\Zbet}{Z_{\beta}}
\newcommand{\Zgam}{Z_{\gamma}}
\newcommand{\Amu}{A_{\mu}}\newcommand{\Anu}{A_{\nu}}
\newcommand{\Zmu}{Z_{\mu}}\newcommand{\Znu}{Z_{\nu}}

\newcommand{\deltae}{\Delta_{\epsilon}}

\begin{document}
\baselineskip=20pt

\pagenumbering{arabic}

\vspace{1.0cm}
\begin{flushright}
LU-ITP 2002/001
\end{flushright}

\begin{center}
{\Large\sf One loop renormalizability of spontaneously broken
gauge theory with a product of gauge groups on noncommutative spacetime:
the $U(1)\times U(1)$ case}\\[10pt]
\vspace{.5 cm}

{Yi Liao}
\vspace{1.0ex}

{\small Institut f\"ur Theoretische Physik, Universit\"at Leipzig,
\\
Augustusplatz 10/11, D-04109 Leipzig, Germany\\}

\vspace{2.0ex}

{\bf Abstract}
\end{center}

A generalization of the standard electroweak model to noncommutative 
spacetime would involve a product gauge group which is 
spontaneously broken. Gauge interactions in terms of physical gauge 
bosons are canonical with respect to massless gauge bosons as 
required by the exact gauge symmetry, but not so with respect to 
massive ones; and furthermore they are generally asymmetric in the 
two sets of gauge bosons. On noncommutative spacetime this already 
occurs for the simplest model of $U(1)\times U(1)$.
We examine whether the above feature in gauge interactions can be 
perturbatively maintained in this model. We show by a complete one 
loop analysis that all ultraviolet divergences are removable with 
a few renormalization constants in a way consistent with the above 
structure.

\begin{flushleft}
PACS: 12.60.-i, 02.40.Gh, 11.10.Gh, 11.15.Ex 

Keywords: noncommutative field theory, spontaneous symmetry breaking, 
gauge theory, renormalization

\end{flushleft}

\newpage
\begin{center}
{\bf 1. Introduction}
\end{center}

The simplest noncommutative (NC) spacetime is the one in which coordinates 
$\hat{x}$ satisfy the algebra, 
$[\hat{x}_{\mu},\hat{x}_{\nu}]=i\theta_{\mu\nu}$, 
where $\theta_{\mu\nu}$ is a real, antisymmetric, constant $n\times n$ 
matrix in $n$ dimensions. A possible way to formulate field theory on this 
NC spacetime is through the Moyal-Weyl correspondence. One starts with the 
Weyl definition of function on NC spacetime by the Fourier transform,
\begin{equation}
\hat{f}(\hat{x})=
\frac{1}{(2\pi)^{n/2}}
\int d^nk~e^{ik_{\mu}\hat{x}^{\mu}}
\tilde{f}(k),
\end{equation}
where $\tilde{f}(k)$ also defines a function $f(x)$ on the ordinary 
commutative spacetime. This relationship is shared by the algebras of 
functions on the two spacetimes respectively if the ordinary product of 
functions on commutative spacetime is replaced by the following star 
product,
\begin{equation}
(f_1\star f_2)(x)=\left[\exp\left(\frac{i}{2}
\theta^{\mu\nu}\partial^x_{\mu}\partial^y_{\nu}\right)
f_1(x)f_2(y)\right]_{y=x}.
\end{equation}
It is in this sense that one may study NC field theory by studying its 
counterpart on commutative spacetime where the ordinary product of 
functions is replaced by the starred one $\cite{review}$.

Field theories on NC spacetime have some salient features that are in 
contrast with ordinary theories and remain to be better understood; 
for example, the causality and unitarity problem $\cite{unitarity}$ 
for time-space noncommutativity and the ultraviolet-infrared(UV-IR) 
mixing $\cite{mixing}$. 
Furthermore, it would be natural to ask whether it is possible to 
generalize gauge interactions to NC spacetime. 
An important ingredient in establishing the viability of the 
generalizaion as a quantum theory is its renormalizability. 
This is a task that has to be fulfilled before one can build up any 
realistic models. It is the purpose of this work to continue the 
pursue in this direction, especially towards constructing realistic 
models for electroweak interactions. Our known results in this aspect 
are mainly based on explicit analyses and a general proof for 
(non)renormalizability of gauge theory on NC spacetime is still 
lacking $\cite{allorders}$. 
This occurs due essentially to the highly nonlocal character 
of NC field theory. The renormalizability of the exact $U(1)$ 
$\cite{u1}$ and $U(N)$ $\cite{un}$ gauge theories has been 
established to one loop order, and that of the real $\phi^4$ 
theory $\cite{phi4}$ to two loops. The situation in spontaneously 
broken gauge theories is more subtle, considering the problems 
already met with spontaneously broken global symmetries 
$\cite{campbell}$. The cases for the broken $U(1)$ $\cite{petriello}$ 
and $U(2)$ $\cite{liao}$ theories have been examined,
both with an affirmative answer. And it would be plausible to expect 
that the latter result also applies to the $U(N)$ ($N>2$) case.

In this work we extend the study of spontaneously broken gauge theories 
on NC spacetime to those with a product of groups. Our basic 
considerations are as follows. 
In a spontaneously broken gauge theory with a single group, the gauge 
couplings of unbroken and broken gauge interactions are the same; and 
the gauge boson masses are also fixed by the group structure. For 
example, for $U(N)$ broken down to $U(N-1)$ by a scalar field in the 
fundamental representation, all $N-1$ pairs of charged gauge bosons ($W$) 
have the same mass which is related to that of the single neutral gauge 
boson ($Z$) by $m_W=m_Z/\sqrt{2}$. This is indeed some distance 
to our goal of constructing realistic electroweak models. It might be 
that for this purpose we have to consider the case with a product of 
groups so that we can have more space for tuning couplings and masses. 
There is a new feature in this case that does not appear for a single 
gauge group, namely the interactions among physical gauge bosons which 
are the mixtures of states originally associated to different group factors. 
Since only some combined part of symmetries is left unbroken, these 
interactions are usually not in a canonical form as dictated by a gauge 
symmetry but have diverse though related coefficients. It is not clear 
whether these relations can still be consistently maintained by 
renormalization at the quantum level on NC spacetime. Furthermore, there 
are not many choices for possible products of groups due to restrictions 
on generalized gauge invariance on NC spacetime in the approach using 
the Moyal-Weyl correspondence. First, only the $U(N)$ group is closed 
under generalized gauge transformations $\cite{closure}$. This explains 
the mass relation mentioned above since there is no freedom even 
for the $U(1)$ part of the group once the kinetic terms are canonically 
normalized. Actually there is no consistent way to separate the $U(N)$ 
group into the group factors of $SU(N)$ and $U(1)$ since the latter 
are always mixed up by generalized gauge transformations. Thus we may 
restrict to the product of $U(N)$ factors. Second, a given matter 
multiplet can have at most two nontrivial representations under two 
of the group factors $\cite{nogo}$. This arises because it is not 
well-defined to transform under more than two group factors due to the 
noncommutativity of the star product. For the purpose of studying 
spontaneous symmetry breaking it would be general enough to consider 
the model of $U(N_1)\times U(N_2)$ with scalars in the 
(anti-)fundamental representations. As a first step of the efforts, we 
shall be less ambitious in this work and consider the simplest case of 
$U(1)_{Y_1}\times U(1)_{Y_2}\to U(1)_Q$ which is nontrivial as compared 
to the one on commutative spacetime because of NC self-interactions. 
We shall see that only the gauge interactions corresponding to $U(1)_Q$ 
are in a canonical form while those of the massive gauge boson are not, 
and that their mixing interactions are asymmetric though related by the 
ratio of couplings. We shall check whether all of this can be maintained 
at one loop level. With only one massive gauge boson we cannot discuss 
the mass relation and its renormalization. But with the positive 
results achieved thus far and in this work it is tempting to expect that 
the same answer would be applicable to the much more complicated case of 
$U(N)\times U(1)$ or even $U(N_1)\times U(N_2)$.

An alternative formalism of NC field theory $\cite{wess}\cite{grosse}$ 
is based on the Seiberg-Witten map $\cite{map}$ which relates NC and 
commutative gauge fields and is solved by a series expansion in $\theta$. 
While it is more flexible to gauge groups and representations, it 
is not clear how to handle with increasingly higher dimension 
operators as one goes to higher orders in
couplings and $\theta$. We shall follow below the naive approach using 
the star product though we are aware of the potential jeopardy at higher 
orders caused by the UV-IR mixing. 

The paper is organized as follows. In the next section we first write 
down the model and emphasize its difference to the commutative case, and 
then introduce the renormalization constants. We demonstrate its one loop 
renormalizability in section 3 by a complete analysis of all 1PI Green's 
functions which may be divergent by power counting. We conclude with the 
last section. We show in the appendices the Feynman rules and 
counterterms of the model and the Feynman diagrams for the 1PI four 
point Green's functions computed in the text.

\begin{center}
{\bf 2. The model}
\end{center}

{\it 2.1 Classical Lagrangian}

We assume that there are two gauge fields $G_{i\mu}$ ($i=1,2$) 
corresponding to the two groups $U(1)_{Y_i}$ with 
respect to both of which the complex scalar field $\Phi$ is charged. 
The generalized, starred gauge transformations are 
\begin{equation}
\begin{array}{rcccl}
G_{i\mu}&\to & G^{\prime}_{i\mu}&=&U_i\star G_{i\mu}\star U_i^{-1}
+ig_i^{-1}U_i\star\partial_{\mu}U_i^{-1},\\
\Phi &\to & \Phi^{\prime}&=&U_1\star\Phi\star U_2^{-1},
\end{array}
\end{equation}
where $U_i=\exp[ig_i\eta_i(x)]_{\star}$, and $g_i$ are gauge couplings. 
Note that the transformation rule for $\Phi$ is unique up to 
interchanging the roles of the two group factors. This arises because 
of the following observation $\cite{nogo}$. Although the two 
symmetries are commutative as global and internal ones, they are not 
so as position-dependent ones due to noncommutativity of the star 
product of ordinary functions. It would be unclear how to do group 
multiplication if we assigned a transformation rule like, e.g., 
$\Phi\to\Phi^{\prime}=U_1\star U_2\star\Phi$.
The classical action invariant under the above transformations 
is constructed from the following Lagrangian,
\begin{equation}
\begin{array}{rcl}
{\cal L}_{\rm class}&=&\displaystyle
-\frac{1}{4}G_{1\mu\nu}G_1^{~\mu\nu}
-\frac{1}{4}G_{2\mu\nu}G_2^{~\mu\nu}\\
&&\displaystyle
+(D_{\mu}\Phi)^{\dagger}D^{\mu}\Phi
+\mu^2\Phi^{\dagger}\Phi-\lambda\Phi^{\dagger}\Phi\Phi^{\dagger}\Phi,
\end{array}
\end{equation}
where we have suppressed the star notation for brevity, and 
\begin{equation}
\begin{array}{rcl}
G_{i\mu\nu}&=&\partial_{\mu}G_{i\nu}-\partial_{\nu}G_{i\mu}
-ig_i[G_{i\mu},G_{i\nu}],\\ 
D_{\mu}\Phi&=&\partial_{\mu}\Phi-ig_1G_{1\mu}\Phi+ig_2\Phi G_{2\mu},
\end{array}
\end{equation}
with $[A,B]\equiv A\star B-B\star A$.

The spontaneous symmetry breaking is triggered by the non-vanishing 
scalar VEV, assuming $\mu^2,\lambda>0$, 
\begin{equation}
\Phi=\phi+\phi_0,~\phi=(\sigma+i\pi)/\sqrt{2},~\phi_0=v/\sqrt{2},
\end{equation}
with $v=\sqrt{\mu^2/\lambda}$. The $\sigma$ field is the physical 
Higgs boson with mass $m_{\sigma}=\sqrt{2\lambda v^2}$ and the 
$\pi$ field is the would-be Goldstone boson. The physical gauge 
bosons are the massless photon $A$ corresponding to the unbroken 
$U(1)_Q$ and the massive $Z$ with mass $m_Z=gv$, where
\begin{equation}
\begin{array}{rcl}
\left(\begin{array}{c}Z\\A
      \end{array}\right)&=&
\left(\begin{array}{lr}c&-s\\s&c
      \end{array}\right)
\left(\begin{array}{c}G_1\\G_2
      \end{array}\right),\\
g&=&\sqrt{g_1^2+g_2^2},~c=g_1/g,~s=g_2/g.       
\end{array}
\label{eq_mix}
\end{equation}
In terms of the above fields, ${\cal L}_{\rm class}$ is expanded 
as a sum of the pure gauge terms and those involving the scalar
fields. The first ones can be cast into the following form,
\begin{equation}
\begin{array}{rcl}
{\cal L}_{2G}&=&\displaystyle
-\frac{1}{4}Z_1Z_1+\frac{1}{2}m_Z^2Z_{\mu}Z^{\mu}
-\frac{1}{4}A_1A_1, \\
{\cal L}_{3G}&=&\displaystyle
\frac{1}{2}ig\left[(c^2-s^2)Z_1Z_2+cs(A_1A_2
+A_1Z_2+Z_1M)\right],\\
{\cal L}_{4G}&=&\displaystyle
\frac{1}{4}g^2\left[(c^6+s^6)Z_2Z_2+c^2s^2(A_2A_2+2Z_2A_2
+MM)+2cs(c^2-s^2)Z_2M\right],
\end{array}
\end{equation}
where we have freely used the property of the star product, 
$\int d^nx fg=\int d^nx gf$, to organize terms, and the following 
notations for brevity,
\begin{equation}
\begin{array}{rclrcl}
Z_1&=&\partial_{\mu}Z_{\nu}-\partial_{\nu}Z_{\mu},&
Z_2&=&[Z_{\mu},Z_{\nu}],\\
A_1&=&\partial_{\mu}A_{\nu}-\partial_{\nu}A_{\mu},&
A_2&=&[A_{\mu},A_{\nu}],\\
M&=&[Z_{\mu},A_{\nu}]-[Z_{\nu},A_{\mu}].&&&
\end{array}
\end{equation}
The terms involving scalar fields are
\begin{equation}
\begin{array}{rcl}
{\cal L}_{\phi G}&=&\displaystyle
-m_ZZ_{\mu}\partial^{\mu}\pi,\\
{\cal L}_{2\phi}&=&\displaystyle
 \frac{1}{2}(\partial_{\mu}\sigma)^2
-\frac{1}{2}m^2_{\sigma}\sigma^2
+\frac{1}{2}(\partial_{\mu}\pi)^2,\\
{\cal L}_{3\phi}&=&\displaystyle
-\lambda v\sigma (\sigma^2+\pi^2),\\
{\cal L}_{4\phi}&=&\displaystyle
-\lambda\left(\frac{1}{4}(\sigma^4+\pi^4)
+\sigma^2\pi^2 -\frac{1}{2}\sigma\pi\sigma\pi\right),\\
{\cal L}_{G2\phi}&=&\displaystyle
+igcsA^{\mu}\left(
 (\partial_{\mu}\sigma\sigma-\sigma\partial_{\mu}\sigma)
+(\partial_{\mu}\pi\pi-\pi\partial_{\mu}\pi)\right)\\
&&\displaystyle
+\frac{1}{2}ig(c^2-s^2)Z^{\mu}\left(
 (\partial_{\mu}\sigma\sigma-\sigma\partial_{\mu}\sigma)
+(\partial_{\mu}\pi\pi-\pi\partial_{\mu}\pi)\right)\\
&&\displaystyle
+\frac{1}{2}gZ^{\mu}\left(
 \pi\partial_{\mu}\sigma-\partial_{\mu}\pi\sigma
+\partial_{\mu}\sigma\pi-\sigma\partial_{\mu}\pi\right),\\
{\cal L}_{\phi 2G}&=&\displaystyle
gm_Z\left(ics[Z_{\mu},A^{\mu}]\pi+Z_{\mu}Z^{\mu}\sigma\right),\\
{\cal L}_{2G2\phi}&=&\displaystyle
g^2\left(\frac{1}{2}(c^4+s^4)\Zmu Z^{\mu}(\sigma^2+\pi^2)
+c^2s^2(\Zmu\sigma Z^{\mu}\sigma+\Zmu\pi Z^{\mu}\pi)\right)\\
&&\displaystyle
+g^2c^2s^2\left(\Amu A^{\mu}(\sigma^2+\pi^2)
-(\Amu\sigma A^{\mu}\sigma+\Amu\pi A^{\mu}\pi)\right)\\
&&\displaystyle
+\frac{1}{2}ig^2(c^2-s^2)\Zmu Z^{\mu}[\pi,\sigma]\\
&&\displaystyle
+g^2cs(c^2-s^2)\left(
\frac{1}{2}(\sigma^2+\pi^2)\{\Amu,Z^{\mu}\}
-(\Amu\sigma Z^{\mu}\sigma+\Amu\pi Z^{\mu}\pi)\right)\\
&&\displaystyle
+ig^2cs\left(
\frac{1}{2}[\pi,\sigma]\{\Amu,Z^{\mu}\}
-(\Amu\sigma Z^{\mu}\pi-\Amu\pi Z^{\mu}\sigma)\right).
\end{array}
\label{eq_can}
\end{equation}

Let us make a few remarks on the above classical Lagrangian. 
On commutative spacetime, it would degenerate trivially into 
the Abelian Higgs model plus a non-interacting pure and exact 
$U(1)$ sector. On NC spacetime, however, because of the 
additional quadratic term in the $U(1)$ field strength and the 
noncommutativity of interactions, the exact $U(1)$ sector not only 
self-interacts but also communicates with the Abelian Higgs 
sector. This makes the theory much more involved and nontrivial. 
The self-interations of the photon are canonical as required by 
the exact $U(1)$ symmetry, with a gauge coupling of $gcs$. This 
is not the case with the $Z$ boson corresponding to the broken 
symmetry. And the mixed interactions are also asymmetric with 
respect to the two bosons. This arises essentially from their 
asymmetric couplings to the scalar field which in turn result in 
the mixing between them as shown in eq. $(\ref{eq_mix})$. On the other hand, 
though asymmetric, all of these interactions are related by the 
only two available gauge couplings. It is thus interesting to 
check whether these relations can be consistent with the 
removal of UV divergences at higher orders and thus be 
possibly maintained in perturbation theory.

{\it 2.2 Gauge fixing and ghost terms}

The procedure of gauge fixing may be generalized directly from 
the commutative theory with the ordinary product replaced by the 
starred one,
\begin{equation}
\begin{array}{rcl}
{\cal L}_{\rm g.f.}&=&\displaystyle
-\frac{1}{2\xi_1}f_1f_1-\frac{1}{2\xi_2}f_2f_2,\\
f_1&=&\partial^{\mu}G_{1\mu}+ig_1\xi_1(\phi^{\dagger}\phi_0
-\phi_0^{\dagger}\phi),\\
f_2&=&\partial^{\mu}G_{2\mu}-ig_2\xi_2(\phi^{\dagger}\phi_0
-\phi_0^{\dagger}\phi).
\end{array}
\end{equation}
Denoting the ghost fields as $c_i$ and $\bar{c}_i$ ($i=1,2$) and 
using the BRS transformations, 
\begin{equation}
\begin{array}{rcl}
{\sf s}G_{i\mu}&=&\partial_{\mu}c_i+ig_i[c_i,G_{i\mu}],\\
{\sf s}\phi&=&ig_1c_1\Phi-ig_2\Phi c_2,\\
{\sf s}\phi^{\dagger}&=&ig_2c_2\Phi^{\dagger}
-ig_1\Phi^{\dagger}c_1,\\
{\sf s}c_i&=&ig_ic_ic_i,\\
{\sf s}\bar{c}_i&=&\displaystyle -\frac{1}{\xi_i}f_i,
\end{array}
\end{equation}
the ghost terms are constructed as
\begin{equation}
{\cal L}_{\rm ghost}=-\bar{c}_1{\sf s}f_1-\bar{c}_2{\sf s}f_2,
\end{equation}
where
\begin{equation}
\begin{array}{rcl}
{\sf s}f_1&=&\partial^{\mu}(\partial_{\mu}c_1+ig_1[c_1,G_{1\mu}])
+g_1^2\xi_1(\Phi^{\dagger}c_1\phi_0+\phi_0^{\dagger}c_1\Phi)\\
&&
-g_1g_2\xi_1(c_2\Phi^{\dagger}\phi_0+\phi_0^{\dagger}\Phi c_2),\\
{\sf s}f_2&=&\partial^{\mu}(\partial_{\mu}c_2+ig_2[c_2,G_{2\mu}])
+g_2^2\xi_2(c_2\Phi^{\dagger}\phi_0+\phi_0^{\dagger}\Phi c_2)\\
&&
-g_1g_2\xi_2(\Phi^{\dagger}c_1\phi_0+\phi_0^{\dagger}c_1\Phi).
\end{array}
\end{equation}
Then, ${\sf s}({\cal L}_{\rm g.f.}+{\cal L}_{\rm ghost})=0$ due 
to ${\sf s}^2f_i=0$. 

To avoid unwanted quadratic $A-Z$ mixing in ${\cal L}_{\rm g.f.}$, 
we work below in the simplified version of $\xi_1=\xi_2=\xi$. 
Introducing the diagonalized ghosts corresponding to the gauge 
bosons $A$ and $Z$,
\begin{equation}
\begin{array}{rcl}
\left(\begin{array}{c}c_Z\\c_A
      \end{array}\right)&=&
\left(\begin{array}{lr}c&-s\\s&c
      \end{array}\right)
\left(\begin{array}{c}c_1\\c_2
      \end{array}\right),\\
\\      
\left(\begin{array}{c}\bar{c}_Z\\ \bar{c}_A
      \end{array}\right)&=&
\left(\begin{array}{lr}c&-s\\s&c
      \end{array}\right)
\left(\begin{array}{c}\bar{c}_1\\ \bar{c}_2
      \end{array}\right),
\end{array}
\label{eq_gh}
\end{equation}
we obtain,
\begin{equation}
\begin{array}{rcl}
{\cal L}_{\rm g.f.}&=&\displaystyle
-\frac{1}{2\xi}\left((\partial^{\mu}Z_{\mu})^2
+(\partial^{\mu}A_{\mu})^2\right)-m_Z\pi\partial^{\mu}Z_{\mu}
-\frac{1}{2}\xi m_Z^2\pi^2,\\
{\cal L}_{\rm ghost}&=&\displaystyle
{\cal L}_{c\bar{c}}+{\cal L}_{\phi c\bar{c}}
+{\cal L}_{Gc\bar{c}},
\end{array}
\end{equation}
where
\begin{equation}
\begin{array}{rcl}
{\cal L}_{c\bar{c}}&=&
-\bar{c}_A\partial^2c_A-\bar{c}_Z(\partial^2+\xi m_Z^2)c_Z,
\\
{\cal L}_{\phi c\bar{c}}&=&\displaystyle
-\frac{1}{2}\xi g^2v\bar{c}_Z\left(
\{c_Z,\sigma\}+(c^2-s^2)i[c_Z,\pi]+2csi[c_A,\pi]\right),\\ 

{\cal L}_{G c\bar{c}}&=&\displaystyle
+igcs\left(\partial^{\mu}\bar{c}_A\left([c_A,A_{\mu}]+[c_Z,Z_{\mu}]\right)
+\partial^{\mu}\bar{c}_Z\left([c_Z,A_{\mu}]+[c_A,Z_{\mu}]\right)\right)\\
&&\displaystyle
+ig(c^2-s^2)\partial^{\mu}\bar{c}_Z[c_Z,Z_{\mu}].
\end{array}
\end{equation}
Note that the $Z\pi$ mixing term in ${\cal L}_{\rm g.f.}$ is 
cancelled by ${\cal L}_{\phi G}$ in eq. $(\ref{eq_can})$. The complete 
Feynman rules are collected in Appendix A.

{\it 2.3 Renormalization constants and counterterms}

Now we introduce renormalization constants for the bare 
quantities. It turns out that in the gauge sector it is 
convenient to introduce renormalization constants for the 
original gauge fields. We have,
\begin{equation}
\begin{array}{rclrcl}
(G_{i\mu})_{\rm B}&=&Z^{1/2}_{G_i}G_{i\mu},&
(\phi)_{\rm B}&=&Z^{1/2}_{\phi}\phi,\\
\displaystyle (g_i)_{\rm B}&=&Z^{-1/2}_{G_i}Z_{g_i}g_i,&
(\lambda)_{\rm B}&=&Z^{-2}_{\phi}Z_{\lambda}\lambda,\\
(\mu^2)_{\rm B}&=&\displaystyle 
Z^{-1}_{\phi}\mu^2\left(
1+\frac{\delta\mu^2}{\mu^2}\right),&
(v)_{\rm B}&=&\displaystyle 
Z^{1/2}_{\phi}v\left(1+\frac{\delta v}{v}\right).
\end{array}
\label{eq_ren1}
\end{equation}
The redundant constant $\delta v$ in the scalar sector will 
be determined by demanding vanishing $\sigma$ tadpole at 
higher orders.

Since there is mixing between gauge bosons, there are two 
equivalent ways to proceed when separating counterterms from 
the bare Lagrangian. One way is to start with eq. $(\ref{eq_ren1})$ and 
define the renormalized $A$ and $Z$ fields in terms of the 
renormalized $c$ and $s$ through eq. $(\ref{eq_mix})$. This proves to 
be convenient for the gauge sector. The alternative way is 
to consider eq. $(\ref{eq_mix})$ as a bare relation and introduce 
counterterms for the bare $c$ and $s$ which are in turn 
determined by $\delta Z_{G_i}$ and $\delta Z_{g_i}$. This 
turns out to be better for organizing the counterterms in 
the gauge-scalar sector. In what follows, our $c$ and $s$ are 
always meant to be renormalized quantities when this 
differentiation is necessary. 

For the gauge fixing and ghost part, the procedure is parallel 
to that in Ref. $\cite{liao}$ though slightly more complicated. 
We choose the quantities appearing in the gauge fixing functions 
$f_i$ to be already renormalized, consider the BRS transformation 
of the renormalized fields, and then introduce the renormalization 
constants for the ghost fields,
\begin{equation}
(c_i)_{\rm B}=Z_{c_i}c_i.
\label{eq_ren2}
\end{equation}
The $c_{Z,A}$ fields are again given by eq. $(\ref{eq_gh})$ in terms of 
the renormalized quantities. 

We shall not present the lengthy expressions for the 
counterterms whose Feynamn rules are listed in appendix A. We 
just comment that there are counterterms to vertices which do not 
appear at tree level. This arises from the mixing of the $A,Z$ 
fields and their different renormalization.

\begin{center}
{\bf 3. One loop renormalizability}
\end{center}

In this section we present our one loop results on UV divergences 
and demonstrate explicitly that the model is renormalizable at 
one loop. We consider only diagrams which may be divergent 
by power counting, but exclude exceptional external momentum 
configurations such as $\theta_{\mu\nu}p^{\nu}=0$ which may cause 
the UV-IR mixing. We work for simplicity in the 
$\xi_1=\xi_2=\xi=1$ gauge and use the dimensional regularization 
in $n=4-2\epsilon$ dimensions for the UV divergence.

\begin{table}
\begin{tabular}{l|l|l}
\hline\hline
$\sigma\sigma\sigma,\sigma\pi\pi$&&
$\pi\pi\pi,\pi\sigma\sigma$\\
\hline
$A\sigma\sigma,A\pi\pi$&&\\
$Z\sigma\sigma,Z\pi\pi,Z\sigma\pi$&$A\sigma\pi$&\\
\hline
$ZZ\sigma,ZA\sigma$&$ZA\sigma$&$ZZ\pi,AA\sigma,AA\pi$\\
\hline
$AAA,ZZZ,ZZA$&$ZAA$&\\
\hline
$\sigma c_Z\bar{c}_Z,\pi c_Z\bar{c}_Z,\pi c_A\bar{c}_Z$&
$\sigma c_A\bar{c}_Z$&
$\sigma c_Z\bar{c}_A,\sigma c_A\bar{c}_A,\pi c_Z\bar{c}_A,
\pi c_A\bar{c}_A$\\
\hline
$Ac_A\bar{c}_A,Ac_Z\bar{c}_Z$&
$Zc_A\bar{c}_A$&\\
$Zc_Z\bar{c}_Z,Zc_A\bar{c}_Z,Zc_Z\bar{c}_A$&
$Ac_A\bar{c}_Z,Ac_Z\bar{c}_A$&\\
\hline\hline
$\sigma\sigma\sigma\sigma,\pi\pi\pi\pi,\sigma\sigma\pi\pi$&
&$\sigma\sigma\sigma\pi,\sigma\pi\pi\pi$\\
\hline
$AA\sigma\sigma,AA\pi\pi$&
&\\
$ZZ\sigma\sigma,ZZ\pi\pi,ZZ\sigma\pi$&&\\
$AZ\sigma\sigma,AZ\pi\pi,AZ\sigma\pi$&$AA\sigma\pi$&\\
\hline
$AAAA,ZZZZ,AAZZ,AZZZ$&$AAAZ$&\\
\hline\hline
\end{tabular}
\caption{All possible three- and four-point 1PI Green's 
functions which may be divergent by power counting. Listed 
in the first, second and third columns are respectively 
the vertices with both tree and counterterm contributions, 
with counterterm but without tree contributions, and without 
either.}
\label{tab_1}
\end{table}

We have exhausted all possible one- to four-point 1PI Green's 
functions, but it is unnecessary to list here the lengthy results. 
Instead, we classify in table $\ref{tab_1}$ all possible three- and 
four-point functions for clarity. The tadpole and two-point 
functions are easy to compute and not listed there. The 
computation of up to three-point functions is similar to 
Ref. $\cite{liao}$ and we refer to that reference for details. 
We shall show our calculation of four-point functions by some 
typical examples. But before doing that, we would like to 
mention that there are many cross-checks which guarantee the 
correctness of our results. Generally, there are a large 
number of Green's functions that have to be made finite by 
adjusting the ten renormalization constants in eqs. $(\ref{eq_ren1})$ 
and $(\ref{eq_ren2})$. 
Due to the $A-Z$ mixing there are vertices that do not 
appear at tree level but have counterterms (as shown in the 
second column in the table $\ref{tab_1}$). The cancellation of divergences 
in these vertices serves as a nontrivial consistency check of the result. 
For some vertices, different terms have different couplings and 
momentum-dependent trigonometric functions, and are thus renormalized  
differently. This also serves as a nontrivial check of the calculation. 
Finally, those listed in the last column must 
be finite by themselves if the model is renormalizable.

Let us start with the $\phi\phi\phi\phi$-type vertex. The 
general Feynman diagrams are shown and numbered in 
appendix B. The simplest ones are the vertices 
$\sigma\sigma\sigma\sigma$ and $\pi\pi\pi\pi$. 
We present our result for the vertex 
$\sigma(p_1)\sigma(p_2)\pi(p_3)\pi(p_4)$ which is richer 
in structure, where $p_i$ are the incoming momenta of the 
particles. The UV divergences are found as follows. 
\begin{equation}
\begin{array}{rcl}
(a)_{15}&=&\displaystyle
ig^4\deltae\left\{c_{12}c_{34}\left[\frac{3}{2}(c^2-s^2)^2+2c^2s^2
+\frac{1}{4}(c^2-s^2)^4\right.\right.\\
&&\displaystyle\left.\left. 
+4c^4s^4+2(c^2-s^2)^2c^2s^2+\frac{1}{4}\right]
-c_{13,24}\left[(c^2-s^2)^2+2c^2s^2\right]\right\}\\
&=&\displaystyle 
ig^4\deltae(c^4+s^4)(2c_{12}c_{34}-c_{13,24}),
\end{array}
\end{equation}
with $c_{ij}=\cos(p_i\wedge p_j),s_{ij}=\sin(p_i\wedge p_j),
c_{ij,kl}=\cos(p_i\wedge p_j+p_k\wedge p_l),
s_{ij,kl}=\sin(p_i\wedge p_j+p_k\wedge p_l)$, 
$p\wedge q=1/2\theta_{\mu\nu}p^{\mu}q^{\nu}$ and 
$\deltae=1/(16\pi^2\epsilon)$.
$(x)_n$ means that there are $n$ diagrams contributing to 
the type-$(x)$ diagram shown in appendix B, not counting 
permutations of external identical particles. Note that 
simplified structures as above are usually reached only upon 
summing up permutated diagrams. 
\begin{equation}
\begin{array}{rcl}
(b)_{8}&=&\displaystyle 
i\lambda g^2\deltae\left[(c^2-s^2)^2+4c^2s^2+1\right]
\left[-2c_{12}c_{34}+2(c_{13,24}-c_{12}c_{34})\right]\\
&=&\displaystyle 
i\lambda g^2\deltae 4(c_{13,24}-2c_{12}c_{34}),\\ 
(c)_{12}&=&\displaystyle 
ig^4\deltae\left\{-c_{12}c_{34}\left[
(c^4+s^4)\left(1+(c^2-s^2)^2\right)
+8c^4s^4+4c^2s^2(c^2-s^2)^2\right]\right.\\
&&\displaystyle 
\left.+2(s_{31}s_{24}+s_{41}s_{23})\left[(c^2-s^2)^2+2c^2s^2\right]
\right\}\\
&=&\displaystyle 
ig^4\deltae 2(c^4+s^4)(c_{13,24}-2c_{12}c_{34}),
\end{array}
\end{equation}
where we have used the momentum conservation and the antisymmetry 
of $\theta_{\mu\nu}$ to obtain,
$s_{31}s_{24}+s_{41}s_{23}=c_{13,24}-c_{12}c_{34}$.
\begin{equation}
\begin{array}{rcl}
(d)_{3}&=&\displaystyle 
i\lambda^2\deltae 8(2c_{12}c_{34}-c_{13,24}),\\ 
(e)_{5}&=&\displaystyle 
ig^4\deltae\left\{4c_{12}c_{34}\left[
(c^4+s^4)^2+4c^4s^4+2c^2s^2(c^2-s^2)^2\right]\right.\\
&&\displaystyle\left.+4(s_{13}s_{24}+s_{14}s_{23})
\left[(c^2-s^2)^2+2c^2s^2\right]\right\}\\
&=&\displaystyle
ig^4\deltae 4(c^4+s^4)(2c_{12}c_{34}-c_{13,24}).
\end{array}
\end{equation}
The total UV divergence of the vertex is then,
\begin{equation}
\begin{array}{rcl}
iV^{\sigma\sigma\pi\pi}(p_1,p_2,p_3,p_4)&=&\displaystyle 
i\deltae\left[3g^4(c^4+s^4)-4g^2\lambda+8\lambda^2\right] 
(2c_{12}c_{34}-c_{13,24}).
\end{array}
\end{equation}

The $GG\phi\phi$-type vertices involve the most types of 
diagrams. We illustrate our calculation by the  
$A_{\mu}(p_1)A_{\nu}(p_2)\pi(p_3)\sigma(p_4)$ vertex which 
has no tree level contribution but does have a counterterm. 
\begin{equation}
\begin{array}{rcl}
(a)_{2}&=&\displaystyle 
+ig^4\deltae\gmunu 2(c^2-s^2)c^2s^2~c_{12}s_{34},\\ 
(b)_{2}&=&\displaystyle 
-ig^4\deltae\gmunu\frac{3}{2}(c^2-s^2)c^2s^2~c_{12}s_{34}, 
\end{array}
\end{equation}
where once again the same structure as the counterterm is 
achieved only upon summing over permutated diagrams.
The type-$(c)$ diagram turns out to be finite as the highest 
power of the loop momentum $k$ actually disappears, e.g., 
$k^{\alpha}k^{\beta}k_{\nu}
P_{\alpha\beta\mu}(k+p_3,-k-p_1-p_3,p_1)=O(k^3)$.
There are no type-$(d)$ and -$(j)$ diagrams at all in this case 
since the photon couplings are diagonal in scalar fields. The 
type-$(e)$ diagram is generally complicated because it involves 
two $GGG$ vertices. But in the current case we only have a 
$Z$-loop to compute. 
\begin{equation}
\begin{array}{rcl}
(e)_{1}&=&\displaystyle 
+ig^4\deltae\gmunu 9(c^2-s^2)c^2s^2~c_{12}s_{34},\\
(f)_{2}&=&\displaystyle 
-ig^4\deltae\gmunu 2(c^2-s^2)c^2s^2~c_{12}s_{34},\\
(g)_{2}&=&\displaystyle 
+ig^4\deltae\gmunu \frac{3}{2}(c^2-s^2)c^2s^2~c_{12}s_{34},\\ 
(h)_{4}&=&\displaystyle 
-ig^4\deltae\gmunu 4(c^2-s^2)c^2s^2~c_{12}s_{34},\\
(i)_{4}&=&\displaystyle 
+ig^4\deltae\gmunu 3(c^2-s^2)c^2s^2~c_{12}s_{34},\\
(k)_{1}&=&\displaystyle 
-ig^4\deltae\gmunu 6(c^2-s^2)c^2s^2~c_{12}s_{34},\\
(l)_{2}&=&\displaystyle 
+ig^4\deltae\gmunu 6(c^2-s^2)c^2s^2~c_{12}s_{34}.
\end{array}
\end{equation}
The overall divergence of the vertex is then,
\begin{equation}
\begin{array}{rcl}
iV^{AA\pi\sigma}_{\mu\nu}(p_1,p_2,p_3,p_4)&=&\displaystyle 
+ig^4\deltae\gmunu 8(c^2-s^2)c^2s^2~c_{12}s_{34}.
\end{array}
\end{equation}

The $GGGG$ vertices are the most difficult part of the 
computation due mainly to the complicated momentum dependent 
trigonometric structures. We choose as a typical example 
the $Z_{\mu}(p_1)Z_{\nu}(p_2)A_{\alpha}(p_3)A_{\beta}(p_4)$ 
vertex to show our calculation, which has a rich structure. 
Diagrams of type-$(a),(c),(d)$ and $(f)$ are easy to compute 
with the results:
\begin{equation}
\begin{array}{rcl}
(a)_{7}&=&\displaystyle 
+ig^4\deltae c^2s^2\left[(c^2-s^2)^2+1\right] 
\\
&&\displaystyle 
\times\frac{2}{3}
[g_{\mu\nu}g_{\alpha\beta}
+g_{\mu\alpha}g_{\beta\nu}+g_{\mu\beta}g_{\nu\alpha}]
\left[2c_{12}c_{34}+c_{13,24}\right],\\ 
(c)_{5}&=&\displaystyle 
-ig^4\deltae c^2s^2\left[(c^2-s^2)^2+2c^2s^2\right] 
\\
&&\displaystyle 
\times\frac{1}{6}
[g_{\mu\nu}g_{\alpha\beta}
+g_{\mu\alpha}g_{\beta\nu}+g_{\mu\beta}g_{\nu\alpha}]
\left[2c_{12}c_{34}+c_{13,24}\right],\\ 
(d)_{10}&=&\displaystyle 
-ig^4\deltae c^2s^2\left[(c^2-s^2)^2+1\right] 
\\
&&\displaystyle 
\times 4
[c_{12}c_{34}g_{\mu\nu}g_{\alpha\beta}
+c_{13}c_{24}g_{\mu\alpha}g_{\beta\nu}
+c_{14}c_{23}g_{\mu\beta}g_{\nu\alpha}],\\ 
(f)_{5}&=&\displaystyle 
+ig^4\deltae c^2s^2 2\left\{
2(c^4+s^4)c_{12}c_{34}g_{\mu\nu}g_{\alpha\beta} 
\right.
\\
&&\displaystyle \left.
+\left[(c^2-s^2)^2+1\right](c_{13}c_{24}g_{\mu\alpha}g_{\beta\nu}
+c_{14}c_{23}g_{\mu\beta}g_{\nu\alpha})\right\}.
\end{array}
\end{equation}
Diagrams $(b),(e)$ and $(g)$ involve multiple pure gauge vertices 
and are more complicated. For example, diagram $(b)$ contains a 
product of four triple-vertex $P_{\mu_1\mu_2\mu_3}(p_1,p_2,p_3)$ 
tensors, whose 
highest power term in loop momentum $k$ provides the UV 
divergence. Upon dropping oscillatory phases involving $k$ and 
doing symmetric loop integration, we may use, 
$$\begin{array}{rl}
&P_{\rho\sigma\alpha}(k,-k,0)P^{\sigma}_{~\tau\beta}(k,-k,0)
P^{\tau}_{~\eta\mu}(k,-k,0)P^{\eta\rho}_{~~~\nu}(k,-k,0)\\
\to&
[47(g_{\mu\nu}g_{\alpha\beta}+g_{\mu\beta}g_{\nu\alpha})
+17g_{\mu\alpha}g_{\beta\nu}](k^2)^2/12.
\end{array}
$$
Combining coefficients of $c$ and $s$, the results are 
\begin{equation}
\begin{array}{rcl}
(b)_{5}&=&\displaystyle 
+ig^4\deltae c^2s^2(c^4+s^4)\\ 
&&\displaystyle 
\times\frac{1}{6}\left\{
c_{12,34}[47(g_{\mu\nu}g_{\alpha\beta}+g_{\mu\beta}g_{\nu\alpha})
+17g_{\mu\alpha}g_{\beta\nu}]\right.\\
&&\displaystyle 
~~~+c_{12,43}[47(g_{\mu\nu}g_{\alpha\beta}+g_{\mu\alpha}g_{\beta\nu})
+17g_{\mu\beta}g_{\nu\alpha}]\\
&&\displaystyle \left.
~~~+c_{13,24}[47(g_{\mu\alpha}g_{\beta\nu}+g_{\mu\beta}g_{\nu\alpha})
+17g_{\mu\nu}g_{\alpha\beta}]\right\},\\ 
(e)_{8}&=&\displaystyle 
-ig^4\deltae c^2s^2(c^4+s^4)
\\
&&\displaystyle 
\times 2\left\{
c_{12}c_{34}(13g_{\mu\nu}g_{\alpha\beta}+g_{\mu\beta}g_{\nu\alpha}
+g_{\mu\alpha}g_{\beta\nu})
+9s_{12}s_{34}(g_{\mu\alpha}g_{\beta\nu}-g_{\mu\beta}g_{\nu\alpha})
\right.\\
&&\displaystyle 
~~~+c_{13}c_{24}(13g_{\mu\alpha}g_{\beta\nu}
+g_{\mu\nu}g_{\alpha\beta}+g_{\mu\beta}g_{\nu\alpha})
+9s_{13}s_{24}(g_{\mu\nu}g_{\alpha\beta}-g_{\mu\beta}g_{\nu\alpha})
\\
&&\displaystyle \left.
~~~+c_{14}c_{23}(13g_{\mu\beta}g_{\nu\alpha}
+g_{\mu\alpha}g_{\beta\nu}+g_{\mu\nu}g_{\alpha\beta})
+9s_{14}s_{23}(g_{\mu\nu}g_{\alpha\beta}-g_{\mu\alpha}g_{\beta\nu})
\right\},\\ 
(g)_{4}&=&\displaystyle 
+ig^4\deltae c^2s^2(c^4+s^4)
\\
&&\displaystyle 
\times 2\left\{
c_{12}c_{34}(4g_{\mu\nu}g_{\alpha\beta}+g_{\mu\beta}g_{\nu\alpha}
+g_{\mu\alpha}g_{\beta\nu})
+9s_{12}s_{34}(g_{\mu\alpha}g_{\beta\nu}-g_{\mu\beta}g_{\nu\alpha})
\right.\\
&&\displaystyle 
~~~+2c_{12}c_{34}(5g_{\mu\nu}g_{\alpha\beta}-g_{\mu\alpha}g_{\beta\nu}
-g_{\mu\beta}g_{\nu\alpha})
+6s_{12}s_{34}(g_{\mu\alpha}g_{\beta\nu}-g_{\mu\beta}g_{\nu\alpha})
\\
&&\displaystyle \left.
~~~+c_{13,24}(7g_{\mu\alpha}g_{\beta\nu}+7g_{\mu\beta}g_{\nu\alpha}
-8g_{\mu\nu}g_{\alpha\beta})\right\}. 
\end{array}
\end{equation}
Using momentum conservation and antisymmetry of $\theta_{\mu\nu}$ 
to deduce the following relations,
$$
\begin{array}{rcl}
c_{ik}c_{jl}&=&\displaystyle\frac{1}{2}
[c_{ij}c_{kl}+s_{ij}s_{kl}+c_{ik,jl}],\\
s_{ik}s_{jl}&=&\displaystyle\frac{1}{2}
[c_{ij}c_{kl}+s_{ij}s_{kl}-c_{ik,jl}],\\
c_{ik,jl}&=&c_{il,jk},
\end{array}
$$
where $\{i,j,k,l\}$ is a permutation of $\{1,2,3,4\}$, and 
expressing all trigonometric functions in terms of 
$c_{12}c_{34},s_{12}s_{34}$ and $c_{13,24}$, we obtain the total 
UV divergence for the vertex,
\begin{equation}
\begin{array}{rcl}
iV^{ZZAA}_{\mu\nu\alpha\beta}(p_1,p_2,p_3,p_4)&=&
ig^4\deltae 2c^2s^2(c^4+s^4)\\
&&\times\left\{
\left[c_{13,24}-c_{12}c_{34}\right]
\left[2g_{\mu\nu}g_{\alpha\beta}-g_{\alpha\mu}g_{\beta\nu}
-g_{\alpha\nu}g_{\beta\mu}\right]\right.\\
&&\left.
-3s_{12}s_{34}\left[g_{\alpha\mu}g_{\beta\nu}
-g_{\alpha\nu}g_{\beta\mu}\right]\right\}.
\end{array}
\end{equation}

Our explicit one loop result may be summarized by the 
following set of renormalization constants in the MS scheme, 
\begin{equation}
\begin{array}{rclrcl}
\delta Z_{G_1}&=&\deltae 3g^2c^2,&
\delta Z_{G_2}&=&\deltae 3g^2s^2,\\
\delta Z_{c_1}&=&\deltae g^2c^2,&
\delta Z_{c_2}&=&\deltae g^2s^2,\\
\delta Z_{\phi}&=&\deltae 2g^2,&&&\\
\delta Z_{g_1}&=&-\deltae 2g^2c^2,&
\delta Z_{g_2}&=&-\deltae 2g^2s^2,\\
\lambda\delta Z_{\lambda}&=&\deltae/2\left[
3g^4(c^4+s^4)-4g^2\lambda+8\lambda^2\right],
&&&\\
\lambda\delta\mu^2/\mu^2&=&-\deltae/2\left[
3g^4(1+2c^2s^2)+2g^2\lambda+4\lambda^2\right],
&\delta v/v&=&\deltae g^2.
\end{array}
\end{equation}
The complete calculation shows that the above set is 
sufficient to remove all UV divergences that appear 
at one loop order. We emphasize this is true in the 
gauge sector though gauge interactions are not symmetric 
with respect to the two physical gauge bosons. The 
renormalization constants for the diagonalized fields 
and their mixings are, 
\begin{equation}
\begin{array}{rclrcl}
\delta Z_{A}&=&\deltae 6g^2c^2s^2,&
\delta Z_{Z}&=&\deltae 3g^2(c^4+s^4),\\
\delta Z_{c_A}&=&\deltae 2g^2c^2s^2,&
\delta Z_{c_Z}&=&\deltae g^2(c^4+s^4),\\
\delta Z_{AZ}&=&\deltae 3g^2cs(c^2-s^2),&
\delta Z_{c_Ac_Z}&=&\deltae g^2cs(c^2-s^2). 
\end{array}
\end{equation}
And the counterterm for the gauge boson mass and the 
renormalization constant for the exact $U(1)$ coupling $e=gcs$ 
defined by $(e)_{\rm B}=Z^{-1/2}_A Z_e e$ are 
\begin{equation}
\begin{array}{rclrcl}
\delta m^2_Z&=&m^2_Z\deltae g^2[4-(c^4+s^4)],&
\delta Z_e&=&-\deltae 4e^2. 
\end{array}
\end{equation}

\begin{center}
{\bf 4. Conclusion}
\end{center}

A generalization of the standard electroweak model to NC 
spacetime would involve a product gauge group which is 
spontaneously broken. A criterion to consider this as a 
viable quantum field theory should include its perturbative   
renormalizability. We pointed out that there are two 
features in such a model which do not appear in the case 
of a single gauge group. Firstly, the gauge boson mass 
relation is determined jointly by the group structure and 
the ratio of gauge couplings. This may allow for more space 
for tuning the masses as happens in the standard model. 
Secondly, the gauge interactions of 
massless gauge bosons are canonical as required by exact 
gauge symmetry, but those of massive ones are generally not. 
The mixed interactions between the two sets of gauge bosons 
are also asymmetric though related, even if we start with a 
symmetric arrangement of group factors like $U(N)\times U(N)$. 
It is the purpose of the current work to examine whether these 
features can be consistently maintained at higher orders in 
perturbation theory so that such a model may still be 
renormalizable on NC spacetime. 
Due to technical complications, we have restricted to the 
simplest case of $U(1)_{Y_1}\times U(1)_{Y_2}\to U(1)_Q$ as 
a first step in these efforts. Although the first feature 
mentioned above never appears, the second one can be 
thoroughly explored. We found indeed all UV divergences at 
one loop level can be removed altogether with a few 
renormalization constants. 
Based on this result and those already achieved so far, it 
would be very natural to expect that the same conclusion also 
applies to the more general case with the 
$U(N_1)\times U(N_2)$ gauge group.
Furthermore, while this result is far away from 
demonstrating renormalizability to all orders, it does lend 
support to the viewpoint that it is worthwhile to consider 
seriously building up realistic models of gauge interactions 
on NC spacetime though this seems to be rather difficult.  

{\bf Acknowledgements}

I would like to thank K. Sibold for many helpful discussions 
and for reading the manuscript carefully.

\newpage
\begin{center}
{\bf Appendix A} Feynman rules
\end{center}

We list below the complete Feynman rules for the model with the 
gauge choice of $\xi_1=\xi_2=\xi$. All momenta 
are incoming and shown in the parentheses of the corresponding 
particles.

\underline{Propagators} (momentum $p$):
\vspace{20pt}

\begin{picture}(80,20)(0,6)
\SetOffset(25,10)
\Photon(0,0)(80,0){3}{8}
\Text(5,10)[]{$\Amu$}\Text(75,10)[]{$\Anu$}
\end{picture}\hspace{30pt}
$\displaystyle=\frac{-i}{p^2}
\left[\gmunu-(1-\xi)\frac{p_{\mu}p_{\nu}}{p^2}\right]$

\begin{picture}(80,20)(0,6)
\SetOffset(25,10)
\Photon(0,0)(80,0){3}{8}
\Text(5,10)[]{$\Zmu$}\Text(75,10)[]{$\Znu$}
\end{picture}\hspace{30pt}
$\displaystyle=\frac{-i}{p^2-m^2_Z}
\left[\gmunu-(1-\xi)\frac{p_{\mu}p_{\nu}}{p^2-\xi m^2_Z}\right]$

\begin{picture}(80,20)(0,6)
\SetOffset(25,10)
\DashLine(0,0)(80,0){5}
\Text(5,6)[]{$\sigma$}\Text(75,6)[]{$\sigma$}
\end{picture}\hspace{30pt}
$\displaystyle=\frac{i}{p^2-m^2_{\sigma}}$

\begin{picture}(80,20)(0,6)
\SetOffset(25,10)
\DashLine(0,0)(80,0){5}
\Text(5,6)[]{$\pi$}\Text(75,6)[]{$\pi$}
\end{picture}\hspace{30pt}
$\displaystyle=\frac{i}{p^2-\xi m^2_Z}$

\begin{picture}(80,20)(0,6)
\SetOffset(25,10)
\DashArrowLine(0,0)(40,0){1}\DashArrowLine(80,0)(40,0){1}
\Text(5,6)[]{$c_A$}\Text(75,6)[]{$\bar{c}_A$}
\end{picture}\hspace{30pt}
$\displaystyle=\frac{i}{p^2}$

\begin{picture}(80,20)(0,6)
\SetOffset(25,10)
\DashArrowLine(0,0)(40,0){1}\DashArrowLine(80,0)(40,0){1}
\Text(5,6)[]{$c_Z$}\Text(75,6)[]{$\bar{c}_Z$}
\end{picture}\hspace{30pt}
$\displaystyle=\frac{i}{p^2-\xi m^2_Z}$

\underline{$G\phi\phi$ vertices}:
$$
\begin{array}{rcl}
\Amu\sigma(p_1)\sigma(p_2)&=&\Amu\pi(p_1)\pi(p_2)\\
&=&\displaystyle 2gcs(p_1-p_2)_{\mu}~s_{12} \\
\Zmu\sigma(p_1)\sigma(p_2)&=&\Zmu\pi(p_1)\pi(p_2)\\
&=&\displaystyle g(c^2-s^2)(p_1-p_2)_{\mu}~s_{12}\\
\Zmu\sigma(p_1)\pi(p_2)&=&\displaystyle
g(p_1-p_2)_{\mu}~c_{12}\\
\end{array}
$$
where $s_{ij}=\sin(p_i\wedge p_j)$ and $c_{ij}=\cos(p_i\wedge p_j)$.

\underline{$GG\phi$ vertices}:
$$
\begin{array}{rcl}
\Zmu(p_1)\Znu(p_2)\sigma&=&\displaystyle i2gm_Z\gmunu~c_{12}\\
\Zmu(p_1)\Anu(p_2)\pi&=&\displaystyle i2gcsm_Z\gmunu~s_{12}\\
\end{array}
$$

\underline{$GG\phi\phi$ vertices}:
$$
\begin{array}{rcl}
\Zmu(p_1)\Znu(p_2)\sigma(p_3)\sigma(p_4)&=&\displaystyle
\Zmu(p_1)\Znu(p_2)\pi(p_3)\pi(p_4)\\
&=&\displaystyle 
i2g^2\gmunu[(c^4+s^4)~c_{12}c_{34}+2c^2s^2~c_{13,24}]\\
\Zmu(p_1)\Znu(p_2)\pi(p_3)\sigma(p_4)&=&\displaystyle
i2g^2(c^2-s^2)\gmunu~c_{12}s_{34}\\
\Amu(p_1)\Anu(p_2)\sigma(p_3)\sigma(p_4)&=&\displaystyle
\Amu(p_1)\Anu(p_2)\pi(p_3)\pi(p_4)\\
&=&\displaystyle i4g^2c^2s^2\gmunu[c_{12}c_{34}-c_{13,24}]\\
\Amu(p_1)\Znu(p_2)\sigma(p_3)\sigma(p_4)&=&\displaystyle
\Amu(p_1)\Znu(p_2)\pi(p_3)\pi(p_4)\\
&=&\displaystyle i2g^2cs(c^2-s^2)\gmunu[c_{12}c_{34}-c_{13,24}]\\
\Amu(p_1)\Znu(p_2)\pi(p_3)\sigma(p_4)&=&\displaystyle 
i2g^2cs\gmunu[c_{12}s_{34}+s_{13,24}]\\
\end{array}
$$
where $s_{ij,kl}=\sin(p_i\wedge p_j+p_k\wedge p_l)$
and $c_{ij,kl}=\cos(p_i\wedge p_j+p_k\wedge p_l)$.

\underline{$GGG$ vertices}:
$$
\begin{array}{rcl}
\Aalf(p_1)\Abet(p_2)\Agam(p_3) &=&\displaystyle
-2gcs~s_{12}~P_{\alpha\beta\gamma}(p_1,p_2,p_3)\\
\Zalf(p_1)\Zbet(p_2)\Zgam(p_3) &=&\displaystyle
-2g(c^2-s^2)~s_{12}~P_{\alpha\beta\gamma}(p_1,p_2,p_3)\\
\Zalf(p_1)\Zbet(p_2)\Agam(p_3) &=&\displaystyle
-2gcs~s_{12}~P_{\alpha\beta\gamma}(p_1,p_2,p_3)\\
\end{array}
$$
where
$$
P_{\alpha\beta\gamma}(p_1,p_2,p_3)=(p_1-p_2)_{\gamma}g_{\alpha\beta}+
(p_2-p_3)_{\alpha}g_{\beta\gamma}+(p_3-p_1)_{\beta}g_{\gamma\alpha}.
$$
Some simple properties of it are useful:
$$
\begin{array}{l}
P_{\alpha\beta\gamma}(p_1,p_2,p_3)
=-P_{\alpha\gamma\beta}(p_1,p_3,p_2)\\
=-P_{\beta\alpha\gamma}(p_2,p_1,p_3)
=-P_{\gamma\beta\alpha}(p_3,p_2,p_1),\\
P_{\alpha\beta\gamma}(p_1,p_2,p_3)+P_{\gamma\alpha\beta}(p_1,p_2,p_3)
=P_{\beta\alpha\gamma}(p_1,p_3,p_2).
\end{array}
$$

\underline{$GGGG$ vertices}:
$$
\begin{array}{l}
A_{\mu_1}(p_1)A_{\mu_2}(p_2)A_{\mu_3}(p_3)A_{\mu_4}(p_4)\\
=\displaystyle -i4g^2c^2s^2
[g^A_{\mu_1\mu_2,\mu_3\mu_4}~s_{12}s_{34}
+g^A_{\mu_3\mu_1,\mu_2\mu_4}~s_{31}s_{24}
+g^A_{\mu_2\mu_3,\mu_1\mu_4}~s_{23}s_{14}]\\
Z_{\mu_1}(p_1)Z_{\mu_2}(p_2)Z_{\mu_3}(p_3)Z_{\mu_4}(p_4)\\
=\displaystyle -i4g^2(c^6+s^6)
[g^A_{\mu_1\mu_2,\mu_3\mu_4}~s_{12}s_{34}
+g^A_{\mu_3\mu_1,\mu_2\mu_4}~s_{31}s_{24}
+g^A_{\mu_2\mu_3,\mu_1\mu_4}~s_{23}s_{14}]\\
Z_{\mu_1}(p_1)Z_{\mu_2}(p_2)A_{\mu_3}(p_3)A_{\mu_4}(p_4)\\
=\displaystyle i2g^2c^2s^2
[g^S_{\mu_1\mu_2,\mu_3\mu_4}(c_{13,24}-c_{12}c_{34})
-3g^A_{\mu_1\mu_2,\mu_3\mu_4}~s_{12}s_{34}]\\
Z_{\mu_1}(p_1)Z_{\mu_2}(p_2)Z_{\mu_3}(p_3)A_{\mu_4}(p_4)\\
=\displaystyle
i4g^2cs(c^2-s^2)[
g_{\mu_4\mu_1}g_{\mu_2\mu_3}(c_{43,12}-c_{41}c_{23})\\
+g_{\mu_4\mu_2}g_{\mu_3\mu_1}(c_{41,23}-c_{42}c_{31})
+g_{\mu_4\mu_3}g_{\mu_1\mu_2}(c_{42,31}-c_{43}c_{12})]
\end{array}
$$
where
$$\begin{array}{rcl}
g^A_{\mu_1\mu_2,\mu_3\mu_4}&=&
g_{\mu_1\mu_3}g_{\mu_2\mu_4}-g_{\mu_1\mu_4}g_{\mu_2\mu_3},\\
g^S_{\mu_1\mu_2,\mu_3\mu_4}&=&
2g_{\mu_1\mu_2}g_{\mu_3\mu_4}
-g_{\mu_1\mu_3}g_{\mu_2\mu_4}-g_{\mu_1\mu_4}g_{\mu_2\mu_3}.
\end{array}
$$

\underline{$\phi\phi\phi$ vertices}:
$$
\begin{array}{rcl}
\sigma\sigma(p_1)\sigma(p_2)&=&\displaystyle
-i6\lambda v~c_{12}\\
\sigma\pi(p_1)\pi(p_2)&=&\displaystyle
-i2\lambda v~c_{12}\\
\end{array}
$$

\underline{$\phi\phi\phi\phi$ vertices}:
$$
\begin{array}{rcl}
\sigma(p_1)\sigma(p_2)\sigma(p_3)\sigma(p_4)&=&
\pi(p_1)\pi(p_2)\pi(p_3)\pi(p_4)\\
&=&-i2\lambda[c_{12}c_{34}+c_{31}c_{24}+c_{23}c_{14}]\\
\sigma(p_1)\sigma(p_2)\pi(p_3)\pi(p_4)&=&
-i2\lambda[2c_{12}c_{34}-c_{13,24}]\\
\end{array}
$$

\underline{$\phi c\bar{c}$ vertices}:
$$
\begin{array}{rcl}
\sigma c_Z(p_1)\bar{c}_Z(p_2)&=&
\displaystyle -i\xi g^2v~c_{21}\\
\pi c_Z(p_1)\bar{c}_Z(p_2)&=&
\displaystyle -i\xi g^2v(c^2-s^2)~s_{21}\\
\pi c_A(p_1)\bar{c}_Z(p_2)&=&
\displaystyle -i\xi g^2v~2cs~s_{21}\\
\end{array}
$$

\underline{$Gc\bar{c}$ vertices}:
$$
\begin{array}{rcl}
\Amu c_A(p_1)\bar{c}_A(p_2)&=&\Amu c_Z(p_1)\bar{c}_Z(p_2)\\
&=&2gcs~p_{2\mu}~s_{21}\\
\Zmu c_Z(p_1)\bar{c}_Z(p_2)&=&
2g(c^2-s^2)p_{2\mu}~s_{21}\\
\Zmu c_A(p_1)\bar{c}_Z(p_2)&=&\Zmu c_Z(p_1)\bar{c}_A(p_2)\\
&=&2gcs~p_{2\mu}~s_{21}\\
\end{array}
$$

Counterterms for self-energies and mixings are listed below. Note the
momentum $p$ is the incoming momentum of the gauge boson in the
$G\phi$ mixing.

\begin{picture}(80,20)(0,6)
\SetOffset(25,10)
\DashLine(0,0)(80,0){5}
\Text(5,6)[]{$\sigma$}\Text(80,0)[]{$\times$}
\end{picture}\hspace{30pt}
$\displaystyle=i\lambda v^3\left[\delta\mu^2/\mu^2
-\delta Z_{\lambda}-2\delta v/v\right]$

\begin{picture}(80,20)(0,6)
\SetOffset(25,10)
\DashLine(0,0)(80,0){5}
\Text(5,6)[]{$\sigma$}\Text(80,6)[]{$\sigma$}\Text(40,0)[]{$\times$}
\end{picture}\hspace{30pt}
$\displaystyle=ip^2\delta Z_{\phi}
-im^2_{\sigma}/2\left[-\delta\mu^2/\mu^2
+3\delta Z_{\lambda}+6\delta v/v\right]$

\begin{picture}(80,20)(0,6)
\SetOffset(25,10)
\DashLine(0,0)(80,0){5}
\Text(5,6)[]{$\pi$}\Text(80,6)[]{$\pi$}\Text(40,0)[]{$\times$}
\end{picture}\hspace{30pt}
$\displaystyle=ip^2\delta Z_{\phi}
-im^2_{\sigma}/2\left[-\delta\mu^2/\mu^2
+\delta Z_{\lambda}+2\delta v/v\right]$

\begin{picture}(80,20)(0,6)
\SetOffset(25,10)
\Photon(0,0)(80,0){3}{8}
\Text(5,10)[]{$\Amu$}\Text(80,10)[]{$\Anu$}\Text(40,0)[]{$\times$}
\end{picture}\hspace{30pt}
$\displaystyle=i(p_{\mu}p_{\nu}-p^2\gmunu)
(s^2\delta Z_{G_1}+c^2\delta Z_{G_2})$

\begin{picture}(80,20)(0,6)
\SetOffset(25,10)
\Photon(0,0)(80,0){3}{8}
\Text(5,10)[]{$\Zmu$}\Text(80,10)[]{$\Znu$}\Text(40,0)[]{$\times$}
\end{picture}\hspace{30pt}
$\displaystyle=i(p_{\mu}p_{\nu}-p^2\gmunu)
(c^2\delta Z_{G_1}+s^2\delta Z_{G_2})\\
\displaystyle~~~~~~~~~~~~~~~~~~~~~~~~~~~~~~~~~~~~~
+i\gmunu m^2_Z\left[2(c^2\delta Z_{g_1}+s^2\delta Z_{g_2})
+2\delta v/v+\delta Z_{\phi}\right]
$

\begin{picture}(80,20)(0,6)
\SetOffset(25,10)
\Photon(0,0)(80,0){3}{8}
\Text(5,10)[]{$\Zmu$}\Text(80,10)[]{$\Anu$}\Text(40,0)[]{$\times$}
\end{picture}\hspace{30pt}
$\displaystyle=i(p_{\mu}p_{\nu}-p^2\gmunu)
cs(\delta Z_{G_1}-\delta Z_{G_2})\\
\displaystyle~~~~~~~~~~~~~~~~~~~~~~~~~~~~~~~~~~~~~
+i\gmunu m^2_Zcs(\delta Z_{g_1}-\delta Z_{g_2})
$

\begin{picture}(80,20)(0,6)
\SetOffset(25,10)
\Photon(0,0)(40,0){3}{4}\DashLine(40,0)(80,0){5}
\Text(5,10)[]{$\Zmu$}\Text(80,10)[]{$\pi$}\Text(40,0)[]{$\times$}
\end{picture}\hspace{30pt}
$\displaystyle=m_Zp_{\mu}\left[(c^2\delta Z_{g_1}+s^2\delta Z_{g_2})
+\delta v/v+\delta Z_{\phi}\right]$

\begin{picture}(80,20)(0,6)
\SetOffset(25,10)
\Photon(0,0)(40,0){3}{4}\DashLine(40,0)(80,0){5}
\Text(5,10)[]{$\Amu$}\Text(80,10)[]{$\pi$}\Text(40,0)[]{$\times$}
\end{picture}\hspace{30pt}
$\displaystyle=m_Zp_{\mu}cs(\delta Z_{g_1}-\delta Z_{g_2})$

\begin{picture}(80,20)(0,6)
\SetOffset(25,10)
\DashArrowLine(0,0)(40,0){1}\DashArrowLine(80,0)(40,0){1}
\Text(5,6)[]{$c_A$}\Text(75,6)[]{$\bar{c}_A$}
\Text(40,0)[]{$\times$}
\end{picture}\hspace{30pt}
$\displaystyle=ip^2(s^2\delta Z_{c_1}+c^2\delta Z_{c_2})$

\begin{picture}(80,20)(0,6)
\SetOffset(25,10)
\DashArrowLine(0,0)(40,0){1}\DashArrowLine(80,0)(40,0){1}
\Text(5,6)[]{$c_Z$}\Text(75,6)[]{$\bar{c}_Z$}
\Text(40,0)[]{$\times$}
\end{picture}\hspace{30pt}
$\displaystyle=ip^2(c^2\delta Z_{c_1}+s^2\delta Z_{c_2})\\
\displaystyle~~~~~~~~~~~~~~~~~~~~~~~~~~~~~~~~~~~~~
-i\xi m^2_Z\left[(c^2\delta Z_{c_1}+s^2\delta Z_{c_2})
+(c^2\delta Z_{g_1}+s^2\delta Z_{g_2})+\delta v/v\right]$

\begin{picture}(80,20)(0,6)
\SetOffset(25,10)
\DashArrowLine(0,0)(40,0){1}\DashArrowLine(80,0)(40,0){1}
\Text(5,6)[]{$c_Z$}\Text(75,6)[]{$\bar{c}_A$}
\Text(40,0)[]{$\times$}
\end{picture}\hspace{30pt}
$\displaystyle=ip^2cs(\delta Z_{c_1}-\delta Z_{c_2})$

\begin{picture}(80,20)(0,6)
\SetOffset(25,10)
\DashArrowLine(0,0)(40,0){1}\DashArrowLine(80,0)(40,0){1}
\Text(5,6)[]{$c_A$}\Text(75,6)[]{$\bar{c}_Z$}
\Text(40,0)[]{$\times$}
\end{picture}\hspace{30pt}
$\displaystyle=ip^2cs(\delta Z_{c_1}-\delta Z_{c_2})\\
\displaystyle~~~~~~~~~~~~~~~~~~~~~~~~~~~~~~~~~~~~~
-i\xi m^2_Zcs\left[(\delta Z_{c_1}-\delta Z_{c_2})
+(\delta Z_{g_1}-\delta Z_{g_2})\right]$

The counterterms for pure scalar vertices are obtained as in $U(2)$ 
theory by attaching appropriate factors to the corresponding 
Feynman rules.
$$
\begin{array}{rl}
\phi\phi\phi:&{\rm tree}\times[\delta Z_{\lambda}+\delta v/v]\\
\phi\phi\phi\phi:&{\rm tree}\times\delta Z_{\lambda}\\
\end{array}
$$
The counterterms for vertices involving gauge bosons become complicated 
due to the $A-Z$ mixing and different renormalization of their fields and 
couplings. There are also counterterms to vertices that do not appear 
at tree level. This is also the case for ghost vertices.
For the vertices appearing already at tree level we use the same notations 
of momenta and indices below for their counterterms.

\underline{$G\phi\phi$ vertices}:
$$
\begin{array}{rl}
A\sigma\sigma,~A\pi\pi:&{\rm tree}\times
[(\delta Z_{g_1}+\delta Z_{g_2})/2+\delta Z_{\phi}]\\
Z\sigma\sigma,~Z\pi\pi:&{\rm tree}\times
[(c^2\delta Z_{g_1}-s^2\delta Z_{g_2})/(c^2-s^2)+\delta Z_{\phi}]\\
Z\pi\sigma:&{\rm tree}\times
[(c^2\delta Z_{g_1}+s^2\delta Z_{g_2})+\delta Z_{\phi}]\\
A_{\mu}\sigma(p_1)\pi(p_2):&gcs(p_1-p_2)_{\mu}~c_{12}
(\delta Z_{g_1}-\delta Z_{g_2})
\end{array}
$$

\underline{$GG\phi$ vertices}:
$$
\begin{array}{rl}
ZZ\sigma:&{\rm tree}\times[2(c^2\delta Z_{g_1}+s^2\delta Z_{g_2})
+\delta Z_{\phi}+\delta v/v]\\
ZA\pi:&{\rm tree}\times[(\delta Z_{g_1}+\delta Z_{g_2})
+\delta Z_{\phi}+\delta v/v]\\
Z_{\mu}(p_1)A_{\nu}(p_2)\sigma:&i2gcsm_Z
\gmunu~c_{12}(\delta Z_{g_1}-\delta Z_{g_2})
\end{array}
$$

\underline{$GG\phi\phi$ vertices}:
$$
\begin{array}{rl}
ZZ\sigma\sigma,~ZZ\pi\pi:&\displaystyle 
i2g^2\gmunu\{
[2(c^4\delta Z_{g_1}+s^4\delta Z_{g_2})+(c^4+s^4)\delta Z_{\phi}]
c_{12}c_{34}\\
&+2c^2s^2[\delta Z_{g_1}+\delta Z_{g_2}+\delta Z_{\phi}]
c_{13,24}\}\\
ZZ\pi\sigma:&{\rm tree}\times[2(c^4\delta Z_{g_1}-s^4\delta Z_{g_2})
/(c^2-s^2)+\delta Z_{\phi}]\\
AA\sigma\sigma,~AA\pi\pi:&
{\rm tree}\times[\delta Z_{g_1}+\delta Z_{g_2}+\delta Z_{\phi}]\\
AZ\sigma\sigma,~AZ\pi\pi:&i2g^2cs\gmunu\{
[2(c^2\delta Z_{g_1}-s^2\delta Z_{g_2})+(c^2-s^2)\delta Z_{\phi}]
c_{12}c_{34}\\
&-(c^2-s^2)[\delta Z_{g_1}+\delta Z_{g_2}+\delta Z_{\phi}]c_{13,24}\}\\
AZ\pi\sigma:&i2g^2cs\gmunu\{
[2(c^2\delta Z_{g_1}+s^2\delta Z_{g_2})+\delta Z_{\phi}]c_{12}s_{34}\\
&+[\delta Z_{g_1}+\delta Z_{g_2}+\delta Z_{\phi}]s_{13,24}\}\\
\Amu(p_1)\Anu(p_2)\pi(p_3)\sigma(p_4):&
i4g^2c^2s^2\gmunu~c_{12}s_{34}(\delta Z_{g_1}-\delta Z_{g_2})\\
\end{array}
$$

\underline{$GGG$ vertices}:
$$
\begin{array}{rl}
AAA:&{\rm tree}\times[(s^2\delta Z_{g_1}+c^2\delta Z_{g_2})
+(s^2\delta Z_{G_1}+c^2\delta Z_{G_2})]\\
ZZZ:&{\rm tree}\times[(c^4\delta Z_{g_1}-s^4\delta Z_{g_2})
+(c^4\delta Z_{G_1}-s^4\delta Z_{G_2})]/(c^2-s^2)\\
ZZA:&{\rm tree}\times[(c^2\delta Z_{g_1}+s^2\delta Z_{g_2})
+(c^2\delta Z_{G_1}+s^2\delta Z_{G_2})]\\
\Aalf(p_1)\Abet(p_2)\Zgam(p_3):&-2gc^2s^2~s_{12}~P_{\alpha\beta\gamma}
(p_1,p_2,p_3)[(\delta Z_{g_1}-\delta Z_{g_2})
+(\delta Z_{G_1}-\delta Z_{G_2})]\\
\end{array}
$$

\underline{$GGGG$ vertices}:
$$
\begin{array}{rl}
AAAA:&{\rm tree}\times[2(s^2\delta Z_{g_1}+c^2\delta Z_{g_2})
+(s^2\delta Z_{G_1}+c^2\delta Z_{G_2})]\\
ZZZZ:&{\rm tree}\times 1/(c^6+s^6)\\
&\times[2(c^6\delta Z_{g_1}+s^6\delta Z_{g_2})
+(c^6\delta Z_{G_1}+s^6\delta Z_{G_2})]\\
ZZAA:&{\rm tree}\times[2(c^2\delta Z_{g_1}+s^2\delta Z_{g_2})
+(c^2\delta Z_{G_1}+s^2\delta Z_{G_2})]\\
ZZZA:&{\rm tree}\times 1/(c^2-s^2)\\
&\times[2(c^4\delta Z_{g_1}-s^4\delta Z_{g_2})
+(c^4\delta Z_{G_1}-s^4\delta Z_{G_2})]\\
A_{\mu_1}(p_1)A_{\mu_2}(p_2)A_{\mu_3}(p_3)Z_{\mu_4}(p_4):&
i4g^2c^3s^3[2(\delta Z_{g_1}-\delta Z_{g_2})
+(\delta Z_{G_1}-\delta Z_{G_2})]\\
&\times[g_{\mu_4\mu_1}g_{\mu_2\mu_3}(c_{43,12}-c_{41}c_{23})\\
&+g_{\mu_4\mu_2}g_{\mu_3\mu_1}(c_{41,23}-c_{42}c_{31})\\
&+g_{\mu_4\mu_3}g_{\mu_1\mu_2}(c_{42,31}-c_{43}c_{12})]
\end{array}
$$

\underline{$\phi c\bar{c}$ vertices}:
$$
\begin{array}{rl}
\sigma c_Z\bar{c}_Z:&{\rm tree}\times[(c^2\delta Z_{g_1}+s^2\delta Z_{g_2})
+(c^2\delta Z_{c_1}+s^2\delta Z_{c_2})]\\
\pi c_Z\bar{c}_Z:&{\rm tree}\times[(c^2\delta Z_{g_1}-s^2\delta Z_{g_2})
+(c^2\delta Z_{c_1}-s^2\delta Z_{c_2})]/(c^2-s^2)\\
\pi c_A\bar{c}_Z:&{\rm tree}\times[(\delta Z_{g_1}+\delta Z_{g_2})
+(\delta Z_{c_1}+\delta Z_{c_2})]/2\\
\sigma c_A(p_1)\bar{c}_Z(p_2):&
-i\xi g^2vcs~c_{21}[(\delta Z_{g_1}-\delta Z_{g_2})
+(\delta Z_{c_1}-\delta Z_{c_2})]
\end{array}
$$

\underline{$Gc\bar{c}$ vertices}:
$$
\begin{array}{rl}
Ac_A\bar{c}_A:&{\rm tree}\times[(s^2\delta Z_{g_1}+c^2\delta Z_{g_2})
+(s^2\delta Z_{c_1}+c^2\delta Z_{c_2})]\\
Ac_Z\bar{c}_Z:&{\rm tree}\times[(c^2\delta Z_{g_1}+s^2\delta Z_{g_2})
+(c^2\delta Z_{c_1}+s^2\delta Z_{c_2})]\\
Zc_Z\bar{c}_Z:&{\rm tree}\times 1/(c^2-s^2)\\
&\times[(c^4\delta Z_{g_1}-s^4\delta Z_{g_2})
+(c^4\delta Z_{c_1}-s^4\delta Z_{c_2})]\\
Zc_A\bar{c}_Z,~Zc_Z\bar{c}_A:&{\rm tree}\times
[(c^2\delta Z_{g_1}+s^2\delta Z_{g_2})
+(c^2\delta Z_{c_1}+s^2\delta Z_{c_2})]\\
\Amu c_A(p_1)\bar{c}_Z(p_2),~\Amu c_Z(p_1)\bar{c}_A(p_2):&
2gc^2s^2p_{2\mu}~s_{21}[(\delta Z_{g_1}-\delta Z_{g_2})
+(\delta Z_{c_1}-\delta Z_{c_2})]\\
\Zmu c_A(p_1)\bar{c}_A(p_2):&
2gc^2s^2p_{2\mu}~s_{21}[(\delta Z_{g_1}-\delta Z_{g_2})
+(\delta Z_{c_1}-\delta Z_{c_2})]\\
\end{array}
$$

\newpage
\begin{center}
{\bf Appendix B} One loop diagrams for 1PI four-point functions
\end{center}

We show below topologically different diagrams in which the wavy, dashed
and dotted lines represent the gauge, scalar and ghost fields respectively.
For a concrete vertex all possible assignments of fields must be
included. Diagrams with an `` f '' are finite by power counting.
The diagrams for two- and three-point functions are similar to those in 
$U(2)$ theory which are shown in Ref. $\cite{liao}$, with the exclusion 
of the charged particles, and will not be repeated here.

$\phi\phi\phi\phi$ vertex:

\begin{center}
\begin{picture}(400,270)(0,0)
\SetOffset(0,180)
\DashLine(20,20)(60,20){5}\DashLine(60,20)(60,60){5}
\DashLine(60,60)(20,60){5}\DashLine(20,60)(20,20){5}
\DashLine(20,0)(20,20){5}\DashLine(60,0)(60,20){5}
\DashLine(60,80)(60,60){5}\DashLine(20,80)(20,60){5}
\Text(40,40)[]{f}
\SetOffset(80,180)
\DashLine(20,20)(60,20){5}\DashLine(60,20)(60,60){5}
\Photon(60,60)(20,60){3}{4}\DashLine(20,60)(20,20){5}
\DashLine(20,0)(20,20){5}\DashLine(60,0)(60,20){5}
\DashLine(60,80)(60,60){5}\DashLine(20,80)(20,60){5}
\Text(40,40)[]{f}
\SetOffset(160,180)
\DashLine(20,20)(60,20){5}\DashLine(60,20)(60,60){5}
\Photon(60,60)(20,60){3}{4}\Photon(20,60)(20,20){3}{4}
\DashLine(20,0)(20,20){5}\DashLine(60,0)(60,20){5}
\DashLine(60,80)(60,60){5}\DashLine(20,80)(20,60){5}
\Text(40,40)[]{f}
\SetOffset(240,180)
\Photon(20,20)(60,20){3}{4}\DashLine(60,20)(60,60){5}
\Photon(60,60)(20,60){3}{4}\DashLine(20,60)(20,20){5}
\DashLine(20,0)(20,20){5}\DashLine(60,0)(60,20){5}
\DashLine(60,80)(60,60){5}\DashLine(20,80)(20,60){5}
\Text(40,40)[]{$(a)$}
\SetOffset(320,180)
\Photon(20,20)(60,20){3}{4}\DashLine(60,20)(60,60){5}
\Photon(60,60)(20,60){3}{4}\Photon(20,60)(20,20){3}{4}
\DashLine(20,0)(20,20){5}\DashLine(60,0)(60,20){5}
\DashLine(60,80)(60,60){5}\DashLine(20,80)(20,60){5}
\Text(40,40)[]{f}

\SetOffset(0,90)
\Photon(20,20)(60,20){3}{4}\Photon(60,20)(60,60){3}{4}
\Photon(60,60)(20,60){3}{4}\Photon(20,60)(20,20){3}{4}
\DashLine(20,0)(20,20){5}\DashLine(60,0)(60,20){5}
\DashLine(60,80)(60,60){5}\DashLine(20,80)(20,60){5}
\Text(40,40)[]{f}
\SetOffset(80,90)
\DashLine(20,20)(60,20){1}\DashLine(60,20)(60,60){1}
\DashLine(60,60)(20,60){1}\DashLine(20,60)(20,20){1}
\DashLine(20,0)(20,20){5}\DashLine(60,0)(60,20){5}
\DashLine(60,80)(60,60){5}\DashLine(20,80)(20,60){5}
\Text(40,40)[]{f}
\SetOffset(160,90)
\DashLine(20,20)(60,20){5}\DashLine(60,20)(40,60){5}
\DashLine(40,60)(20,20){5}
\DashLine(20,0)(20,20){5}\DashLine(60,0)(60,20){5}
\DashLine(20,80)(40,60){5}\DashLine(40,60)(60,80){5}
\Text(40,40)[]{f}
\SetOffset(240,90)
\Photon(20,20)(60,20){3}{4}\DashLine(60,20)(40,60){5}
\DashLine(40,60)(20,20){5}
\DashLine(20,0)(20,20){5}\DashLine(60,0)(60,20){5}
\DashLine(20,80)(40,60){5}\DashLine(40,60)(60,80){5}
\Text(40,40)[]{$(b)$}
\SetOffset(320,90)
\DashLine(20,20)(60,20){5}\Photon(60,20)(40,60){-3}{5}
\Photon(40,60)(20,20){3}{5}
\DashLine(20,0)(20,20){5}\DashLine(60,0)(60,20){5}
\DashLine(20,80)(40,60){5}\DashLine(40,60)(60,80){5}
\Text(40,40)[]{$(c)$}

\SetOffset(0,0)
\Photon(20,20)(60,20){3}{4}\Photon(60,20)(40,60){-3}{5}
\Photon(40,60)(20,20){3}{5}
\DashLine(20,0)(20,20){5}\DashLine(60,0)(60,20){5}
\DashLine(20,80)(40,60){5}\DashLine(40,60)(60,80){5}
\Text(40,40)[]{f}
\SetOffset(80,0)
\DashCArc(40,40)(20,0,360){5}
\DashLine(20,0)(40,20){5}\DashLine(40,20)(60,0){5}
\DashLine(20,80)(40,60){5}\DashLine(40,60)(60,80){5}
\Text(40,40)[]{$(d)$}
\SetOffset(160,0)
\PhotonArc(40,40)(20,0,360){2}{12}
\DashLine(20,0)(40,20){5}\DashLine(40,20)(60,0){5}
\DashLine(20,80)(40,60){5}\DashLine(40,60)(60,80){5}
\Text(40,40)[]{$(e)$}
\end{picture}\\
\end{center}

$\phi\phi GG$ vertex:

\begin{center}
\begin{picture}(400,180)(0,0)
\SetOffset(0,90)
\DashLine(20,20)(60,20){5}\DashLine(60,20)(60,60){5}
\DashLine(60,60)(20,60){5}\DashLine(20,60)(20,20){5}
\Photon(20,0)(20,20){3}{2}\Photon(60,0)(60,20){3}{2}
\DashLine(60,80)(60,60){5}\DashLine(20,80)(20,60){5}
\Text(40,40)[]{f}
\SetOffset(80,90)
\DashLine(20,20)(60,20){5}\DashLine(60,20)(60,60){5}
\Photon(60,60)(20,60){3}{4}\DashLine(20,60)(20,20){5}
\Photon(20,0)(20,20){3}{2}\Photon(60,0)(60,20){3}{2}
\DashLine(60,80)(60,60){5}\DashLine(20,80)(20,60){5}
\Text(40,40)[]{$(a)$}
\SetOffset(160,90)
\DashLine(20,20)(60,20){5}\DashLine(60,20)(60,60){5}
\DashLine(60,60)(20,60){5}\Photon(20,60)(20,20){3}{4}
\Photon(20,0)(20,20){3}{2}\Photon(60,0)(60,20){3}{2}
\DashLine(60,80)(60,60){5}\DashLine(20,80)(20,60){5}
\Text(40,40)[]{f}
\SetOffset(240,90)
\Photon(20,20)(60,20){3}{4}\DashLine(60,20)(60,60){5}
\DashLine(60,60)(20,60){5}\DashLine(20,60)(20,20){5}
\Photon(20,0)(20,20){3}{2}\Photon(60,0)(60,20){3}{2}
\DashLine(60,80)(60,60){5}\DashLine(20,80)(20,60){5}
\Text(40,40)[]{f}
\SetOffset(320,90)
\DashLine(20,20)(60,20){5}\DashLine(60,20)(60,60){5}
\Photon(60,60)(20,60){3}{4}\Photon(20,60)(20,20){3}{4}
\Photon(20,0)(20,20){3}{2}\Photon(60,0)(60,20){3}{2}
\DashLine(60,80)(60,60){5}\DashLine(20,80)(20,60){5}
\Text(40,40)[]{f}

\SetOffset(0,0)
\Photon(20,20)(60,20){3}{4}\DashLine(60,20)(60,60){5}
\Photon(60,60)(20,60){3}{4}\DashLine(20,60)(20,20){5}
\Photon(20,0)(20,20){3}{2}\Photon(60,0)(60,20){3}{2}
\DashLine(60,80)(60,60){5}\DashLine(20,80)(20,60){5}
\Text(40,40)[]{f}
\SetOffset(80,0)
\Photon(20,20)(60,20){3}{4}\DashLine(60,20)(60,60){5}
\DashLine(60,60)(20,60){5}\Photon(20,60)(20,20){3}{4}
\Photon(20,0)(20,20){3}{2}\Photon(60,0)(60,20){3}{2}
\DashLine(60,80)(60,60){5}\DashLine(20,80)(20,60){5}
\Text(40,40)[]{f}
\SetOffset(160,0)
\DashLine(20,20)(60,20){5}\Photon(60,20)(60,60){3}{4}
\DashLine(60,60)(20,60){5}\Photon(20,60)(20,20){3}{4}
\Photon(20,0)(20,20){3}{2}\Photon(60,0)(60,20){3}{2}
\DashLine(60,80)(60,60){5}\DashLine(20,80)(20,60){5}
\Text(40,40)[]{f}
\SetOffset(240,0)
\Photon(20,20)(60,20){3}{4}\Photon(60,20)(60,60){3}{4}
\DashLine(60,60)(20,60){5}\Photon(20,60)(20,20){3}{4}
\Photon(20,0)(20,20){3}{2}\Photon(60,0)(60,20){3}{2}
\DashLine(60,80)(60,60){5}\DashLine(20,80)(20,60){5}
\Text(40,40)[]{$(b)$}
\SetOffset(320,0)
\Photon(20,20)(60,20){3}{4}\Photon(60,20)(60,60){3}{4}
\Photon(60,60)(20,60){3}{4}\DashLine(20,60)(20,20){5}
\Photon(20,0)(20,20){3}{2}\Photon(60,0)(60,20){3}{2}
\DashLine(60,80)(60,60){5}\DashLine(20,80)(20,60){5}
\Text(40,40)[]{f}
\end{picture}\\
\end{center}

\begin{center}
\begin{picture}(400,540)(0,0)
\SetOffset(0,450)
\DashLine(20,20)(60,20){5}\Photon(60,20)(60,60){3}{4}
\Photon(60,60)(20,60){3}{4}\Photon(20,60)(20,20){3}{4}
\Photon(20,0)(20,20){3}{2}\Photon(60,0)(60,20){3}{2}
\DashLine(60,80)(60,60){5}\DashLine(20,80)(20,60){5}
\Text(40,40)[]{f}
\SetOffset(80,450)
\Photon(20,20)(60,20){3}{4}\Photon(60,20)(60,60){3}{4}
\Photon(60,60)(20,60){3}{4}\Photon(20,60)(20,20){3}{4}
\Photon(20,0)(20,20){3}{2}\Photon(60,0)(60,20){3}{2}
\DashLine(60,80)(60,60){5}\DashLine(20,80)(20,60){5}
\Text(40,40)[]{f}
\SetOffset(160,450)
\DashLine(20,20)(60,20){1}\DashLine(60,20)(60,60){1}
\DashLine(60,60)(20,60){1}\DashLine(20,60)(20,20){1}
\Photon(20,0)(20,20){3}{2}\Photon(60,0)(60,20){3}{2}
\DashLine(60,80)(60,60){5}\DashLine(20,80)(20,60){5}
\Text(40,40)[]{f}
\SetOffset(240,450)
\DashLine(20,20)(60,20){5}\DashLine(60,20)(60,60){5}
\DashLine(60,60)(20,60){5}\DashLine(20,60)(20,20){5}
\Photon(20,0)(20,20){3}{2}\DashLine(60,0)(60,20){5}
\Photon(60,80)(60,60){3}{2}\DashLine(20,80)(20,60){5}
\Text(40,40)[]{f}
\SetOffset(320,450)
\DashLine(20,20)(60,20){5}\DashLine(60,20)(60,60){5}
\Photon(60,60)(20,60){3}{4}\DashLine(20,60)(20,20){5}
\Photon(20,0)(20,20){3}{2}\DashLine(60,0)(60,20){5}
\Photon(60,80)(60,60){3}{2}\DashLine(20,80)(20,60){5}
\Text(40,40)[]{f}

\SetOffset(0,360)
\DashLine(20,20)(60,20){5}\DashLine(60,20)(60,60){5}
\Photon(60,60)(20,60){3}{4}\Photon(20,60)(20,20){3}{4}
\Photon(20,0)(20,20){3}{2}\DashLine(60,0)(60,20){5}
\Photon(60,80)(60,60){3}{2}\DashLine(20,80)(20,60){5}
\Text(40,40)[]{f}
\SetOffset(80,360)
\Photon(20,20)(60,20){3}{4}\DashLine(60,20)(60,60){5}
\Photon(60,60)(20,60){3}{4}\DashLine(20,60)(20,20){5}
\Photon(20,0)(20,20){3}{2}\DashLine(60,0)(60,20){5}
\Photon(60,80)(60,60){3}{2}\DashLine(20,80)(20,60){5}
\Text(40,40)[]{f}
\SetOffset(160,360)
\DashLine(20,20)(60,20){5}\Photon(60,20)(60,60){3}{4}
\Photon(60,60)(20,60){3}{4}\DashLine(20,60)(20,20){5}
\Photon(20,0)(20,20){3}{2}\DashLine(60,0)(60,20){5}
\Photon(60,80)(60,60){3}{2}\DashLine(20,80)(20,60){5}
\Text(40,40)[]{$(c)$}
\SetOffset(240,360)
\Photon(20,20)(60,20){3}{4}\Photon(60,20)(60,60){3}{4}
\DashLine(60,60)(20,60){5}\Photon(20,60)(20,20){3}{4}
\Photon(20,0)(20,20){3}{2}\DashLine(60,0)(60,20){5}
\Photon(60,80)(60,60){3}{2}\DashLine(20,80)(20,60){5}
\Text(40,40)[]{f}
\SetOffset(320,360)
\Photon(20,20)(60,20){3}{4}\Photon(60,20)(60,60){3}{4}
\Photon(60,60)(20,60){3}{4}\Photon(20,60)(20,20){3}{4}
\Photon(20,0)(20,20){3}{2}\DashLine(60,0)(60,20){5}
\Photon(60,80)(60,60){3}{2}\DashLine(20,80)(20,60){5}
\Text(40,40)[]{f}

\SetOffset(0,270)
\DashLine(20,20)(60,20){1}\DashLine(60,20)(60,60){1}
\DashLine(60,60)(20,60){1}\DashLine(20,60)(20,20){1}
\Photon(20,0)(20,20){3}{2}\DashLine(60,0)(60,20){5}
\Photon(60,80)(60,60){3}{2}\DashLine(20,80)(20,60){5}
\Text(40,40)[]{f}
\SetOffset(80,270)
\DashLine(20,20)(60,20){5}\DashLine(60,20)(40,60){5}
\DashLine(40,60)(20,20){5}
\Photon(20,0)(20,20){3}{2}\Photon(60,0)(60,20){3}{2}
\DashLine(20,80)(40,60){5}\DashLine(40,60)(60,80){5}
\Text(40,40)[]{$(d)$}
\SetOffset(160,270)
\Photon(20,20)(60,20){3}{4}\DashLine(60,20)(40,60){5}
\DashLine(40,60)(20,20){5}
\Photon(20,0)(20,20){3}{2}\Photon(60,0)(60,20){3}{2}
\DashLine(20,80)(40,60){5}\DashLine(40,60)(60,80){5}
\Text(40,40)[]{f}
\SetOffset(240,270)
\DashLine(20,20)(60,20){5}\Photon(60,20)(40,60){-3}{4}
\Photon(40,60)(20,20){3}{4}
\Photon(20,0)(20,20){3}{2}\Photon(60,0)(60,20){-3}{2}
\DashLine(20,80)(40,60){5}\DashLine(40,60)(60,80){5}
\Text(40,40)[]{f}
\SetOffset(320,270)
\Photon(20,20)(60,20){3}{4}\Photon(60,20)(40,60){-3}{4}
\Photon(40,60)(20,20){3}{4}
\Photon(20,0)(20,20){3}{2}\Photon(60,0)(60,20){-3}{2}
\DashLine(20,80)(40,60){5}\DashLine(40,60)(60,80){5}
\Text(40,40)[]{$(e)$}

\SetOffset(0,180)
\DashLine(20,20)(60,20){5}\DashLine(60,20)(40,60){5}
\DashLine(40,60)(20,20){5}
\DashLine(20,0)(20,20){5}\DashLine(60,0)(60,20){5}
\Photon(20,80)(40,60){3}{3}\Photon(40,60)(60,80){-3}{3}
\Text(40,40)[]{f}
\SetOffset(80,180)
\Photon(20,20)(60,20){3}{4}\DashLine(60,20)(40,60){5}
\DashLine(40,60)(20,20){5}
\DashLine(20,0)(20,20){5}\DashLine(60,0)(60,20){5}
\Photon(20,80)(40,60){3}{3}\Photon(40,60)(60,80){-3}{3}
\Text(40,40)[]{$(f)$}
\SetOffset(160,180)
\DashLine(20,20)(60,20){5}\Photon(60,20)(40,60){-3}{4}
\Photon(40,60)(20,20){3}{4}
\DashLine(20,0)(20,20){5}\DashLine(60,0)(60,20){5}
\Photon(20,80)(40,60){3}{3}\Photon(40,60)(60,80){-3}{3}
\Text(40,40)[]{$(g)$}
\SetOffset(240,180)
\Photon(20,20)(60,20){3}{4}\Photon(60,20)(40,60){-3}{4}
\Photon(40,60)(20,20){3}{4}
\DashLine(20,0)(20,20){5}\DashLine(60,0)(60,20){5}
\Photon(20,80)(40,60){3}{3}\Photon(40,60)(60,80){-3}{3}
\Text(40,40)[]{f}
\SetOffset(320,180)
\DashLine(20,20)(60,20){5}\DashLine(60,20)(40,60){5}
\Photon(40,60)(20,20){-3}{4}
\DashLine(20,0)(20,20){5}\Photon(60,0)(60,20){3}{2}
\DashLine(20,80)(40,60){5}\Photon(40,60)(60,80){-3}{3}
\Text(40,40)[]{$(h)$}

\SetOffset(0,90)
\DashLine(20,20)(60,20){5}\Photon(60,20)(40,60){3}{4}
\DashLine(40,60)(20,20){5}
\DashLine(20,0)(20,20){5}\Photon(60,0)(60,20){3}{2}
\DashLine(20,80)(40,60){5}\Photon(40,60)(60,80){-3}{3}
\Text(40,40)[]{f}
\SetOffset(80,90)
\Photon(20,20)(60,20){-3}{4}\DashLine(60,20)(40,60){5}
\Photon(40,60)(20,20){-3}{4}
\DashLine(20,0)(20,20){5}\Photon(60,0)(60,20){3}{2}
\DashLine(20,80)(40,60){5}\Photon(40,60)(60,80){-3}{3}
\Text(40,40)[]{f}
\SetOffset(160,90)
\Photon(20,20)(60,20){3}{4}\Photon(60,20)(40,60){3}{4}
\DashLine(40,60)(20,20){5}
\DashLine(20,0)(20,20){5}\Photon(60,0)(60,20){3}{2}
\DashLine(20,80)(40,60){5}\Photon(40,60)(60,80){-3}{3}
\Text(40,40)[]{$(i)$}
\SetOffset(240,90)
\DashCArc(40,40)(20,0,360){5}
\Photon(20,0)(40,20){3}{3}\Photon(40,20)(60,0){-3}{3}
\DashLine(20,80)(40,60){5}\DashLine(40,60)(60,80){5}
\Text(40,40)[]{$(j)$}
\SetOffset(320,90)
\PhotonArc(40,40)(20,0,360){2}{12}
\Photon(20,0)(40,20){3}{3}\Photon(40,20)(60,0){-3}{3}
\DashLine(20,80)(40,60){5}\DashLine(40,60)(60,80){5}
\Text(40,40)[]{$(k)$}

\SetOffset(0,0)
\DashCArc(40,40)(20,90,270){5}
\PhotonArc(40,40)(20,-90,90){-2}{6}
\DashLine(20,0)(40,20){5}\Photon(40,20)(60,0){-3}{3}
\DashLine(20,80)(40,60){5}\Photon(40,60)(60,80){-3}{3}
\Text(40,40)[]{$(l)$}
\end{picture}\\
\end{center}

\newpage
$GGGG$ vertex:

\begin{center}
\begin{picture}(400,270)(0,0)
\SetOffset(0,180)
\DashLine(20,20)(60,20){5}\DashLine(60,20)(60,60){5}
\DashLine(60,60)(20,60){5}\DashLine(20,60)(20,20){5}
\Photon(20,0)(20,20){3}{2}\Photon(60,0)(60,20){3}{2}
\Photon(60,80)(60,60){3}{2}\Photon(20,80)(20,60){3}{2}
\Text(40,40)[]{$(a)$}
\SetOffset(80,180)
\DashLine(20,20)(60,20){5}\DashLine(60,20)(60,60){5}
\Photon(60,60)(20,60){3}{4}\DashLine(20,60)(20,20){5}
\Photon(20,0)(20,20){3}{2}\Photon(60,0)(60,20){3}{2}
\Photon(60,80)(60,60){3}{2}\Photon(20,80)(20,60){3}{2}
\Text(40,40)[]{f}
\SetOffset(160,180)
\DashLine(20,20)(60,20){5}\DashLine(60,20)(60,60){5}
\Photon(60,60)(20,60){3}{4}\Photon(20,60)(20,20){3}{4}
\Photon(20,0)(20,20){3}{2}\Photon(60,0)(60,20){3}{2}
\Photon(60,80)(60,60){3}{2}\Photon(20,80)(20,60){3}{2}
\Text(40,40)[]{f}
\SetOffset(240,180)
\Photon(20,20)(60,20){3}{4}\DashLine(60,20)(60,60){5}
\Photon(60,60)(20,60){3}{4}\DashLine(20,60)(20,20){5}
\Photon(20,0)(20,20){3}{2}\Photon(60,0)(60,20){3}{2}
\Photon(60,80)(60,60){3}{2}\Photon(20,80)(20,60){3}{2}
\Text(40,40)[]{f}
\SetOffset(320,180)
\Photon(20,20)(60,20){3}{4}\DashLine(60,20)(60,60){5}
\Photon(60,60)(20,60){3}{4}\Photon(20,60)(20,20){3}{4}
\Photon(20,0)(20,20){3}{2}\Photon(60,0)(60,20){3}{2}
\Photon(60,80)(60,60){3}{2}\Photon(20,80)(20,60){3}{2}
\Text(40,40)[]{f}

\SetOffset(0,90)
\Photon(20,20)(60,20){3}{4}\Photon(60,20)(60,60){3}{4}
\Photon(60,60)(20,60){3}{4}\Photon(20,60)(20,20){3}{4}
\Photon(20,0)(20,20){3}{2}\Photon(60,0)(60,20){3}{2}
\Photon(60,80)(60,60){3}{2}\Photon(20,80)(20,60){3}{2}
\Text(40,40)[]{$(b)$}
\SetOffset(80,90)
\DashLine(20,20)(60,20){1}\DashLine(60,20)(60,60){1}
\DashLine(60,60)(20,60){1}\DashLine(20,60)(20,20){1}
\Photon(20,0)(20,20){3}{2}\Photon(60,0)(60,20){3}{2}
\Photon(60,80)(60,60){3}{2}\Photon(20,80)(20,60){3}{2}
\Text(40,40)[]{$(c)$}
\SetOffset(160,90)
\DashLine(20,20)(60,20){5}\DashLine(60,20)(40,60){5}
\DashLine(40,60)(20,20){5}
\Photon(20,0)(20,20){3}{2}\Photon(60,0)(60,20){3}{2}
\Photon(20,80)(40,60){3}{3}\Photon(40,60)(60,80){-3}{3}
\Text(40,40)[]{$(d)$}
\SetOffset(240,90)
\Photon(20,20)(60,20){3}{4}\DashLine(60,20)(40,60){5}
\DashLine(40,60)(20,20){5}
\Photon(20,0)(20,20){3}{2}\Photon(60,0)(60,20){3}{2}
\Photon(20,80)(40,60){3}{3}\Photon(40,60)(60,80){-3}{3}
\Text(40,40)[]{f}
\SetOffset(320,90)
\DashLine(20,20)(60,20){5}\Photon(60,20)(40,60){-3}{5}
\Photon(40,60)(20,20){3}{5}
\Photon(20,0)(20,20){3}{2}\Photon(60,0)(60,20){3}{2}
\Photon(20,80)(40,60){3}{3}\Photon(40,60)(60,80){-3}{3}
\Text(40,40)[]{f}

\SetOffset(0,0)
\Photon(20,20)(60,20){3}{4}\Photon(60,20)(40,60){-3}{5}
\Photon(40,60)(20,20){3}{5}
\Photon(20,0)(20,20){3}{2}\Photon(60,0)(60,20){3}{2}
\Photon(20,80)(40,60){3}{3}\Photon(40,60)(60,80){-3}{3}
\Text(40,40)[]{$(e)$}
\SetOffset(80,0)
\DashCArc(40,40)(20,0,360){5}
\Photon(20,0)(40,20){3}{3}\Photon(40,20)(60,0){-3}{3}
\Photon(20,80)(40,60){3}{3}\Photon(40,60)(60,80){-3}{3}
\Text(40,40)[]{$(f)$}
\SetOffset(160,0)
\PhotonArc(40,40)(20,0,360){2}{12}
\Photon(20,0)(40,20){3}{3}\Photon(40,20)(60,0){-3}{3}
\Photon(20,80)(40,60){3}{3}\Photon(40,60)(60,80){-3}{3}
\Text(40,40)[]{$(g)$}
\end{picture}\\
\end{center}

\newpage

\end{document}